\documentclass[sigconf,edbt]{acmart-edbt2020}

\def\BibTeX{{\rm B\kern-.05em{\sc i\kern-.025em b}\kern-.08em
    T\kern-.1667em\lower.7ex\hbox{E}\kern-.125emX}}

\usepackage{graphicx}
\usepackage{balance}  
\usepackage[edges]{forest}
\usepackage{tikz}
\usepackage[font=scriptsize, skip=0pt]{subcaption}
\usepackage[linesnumbered,ruled,vlined, noend]{algorithm2e}
\usetikzlibrary{positioning}
\usepackage{textcomp}
\usepackage{multirow}
\usepackage{array}
\usepackage{amsmath}
\usepackage{manfnt}
\usepackage{booktabs}
\usepackage{color, colortbl}
\usepackage[bottom]{footmisc}
\usepackage{tabularx}
\usepackage{pifont}
\usepackage{pgfplots}
\usepgfplotslibrary{groupplots}
\usetikzlibrary{patterns}
\usepackage{enumitem}
\usepackage{filecontents}
\usepackage[title]{appendix}
\usepgfplotslibrary{statistics}

\pgfplotsset{compat=1.16}

\newcommand{\para}[1]{\smallskip\noindent{\bf{#1}}}

\newcommand{\ccube}{\normalfont\mancube}

\setcopyright{rightsretained}

\acmDOI{}

\acmISBN{978-3-89318-084-4}

\acmConference[DOLAP 2021]{23rd International Workshop on Design, Optimization, Languages and Analytical Processing of Big Data (DOLAP)}{March 23, 2021}{Nicosia, Cyprus} 
\acmYear{2021}

\begin{document}
\title{A Foundation for Spatio-Textual-Temporal Cube Analytics (Extended Version)}

\author{Mohsin Iqbal}
\orcid{1234-5678-9012}
\affiliation{%
  \institution{Aalborg University}
}
\email{mohsin@cs.aau.dk}

\author{Matteo Lissandrini}
\orcid{1234-5678-9012}
\affiliation{%
	\institution{Aalborg University}
}
\email{matteo@cs.aau.dk}

\author{Torben Bach Pedersen}
\orcid{1234-5678-9012}
\affiliation{%
	\institution{Aalborg University}
}
\email{tbp@cs.aau.dk}

\renewcommand{\shortauthors}{}

\begin{abstract}
Large amounts of \emph{spatial, textual, and temporal data} are being produced daily.
This is data containing an \emph{unstructured} component (text), a \emph{spatial} component (geographic position), and a \emph{time} component (timestamp).
Therefore, there is a need for a powerful and \emph{general} way of analyzing \emph{spatial, textual, and temporal data together}.
In this paper, we define and formalize the \emph{Spatio-Textual-Temporal Cube} structure to enable \emph{combined} effective and efficient analytical queries over \emph{spatial, textual, and temporal data}.
Our novel data model over \emph{spatio-textual-temporal objects} enables novel \emph{joint and integrated} spatial, textual, and temporal insights that are hard to obtain using existing methods.
Moreover, we introduce the new concept of \emph{spatio-textual-temporal measures} with associated novel spatio-textual-temporal-OLAP operators.
To allow for efficient large-scale analytics, we present a pre-aggregation framework for exact and approximate computation of \emph{spatio-textual-temporal measures}.
Our comprehensive experimental evaluation on a real-world Twitter dataset confirms that our proposed methods reduce query response time by 1-5 orders of magnitude compared to the \emph{No Materialization} baseline and decrease storage cost between 97\% and 99.9\% compared to the \emph{Full Materialization} baseline while adding only a negligible overhead in the Spatio-Textual-Temporal Cube construction time.
Moreover, \textit{approximate computation} achieves an accuracy between 90\% and 100\% while reducing query response time by 3-5 orders of magnitude compared to \emph{No Materialization}.
\end{abstract}

\maketitle

\section{Introduction} \label{sec:intro}
\begin{figure}[t!]
	\centering
	\includegraphics[width=0.3\textwidth]{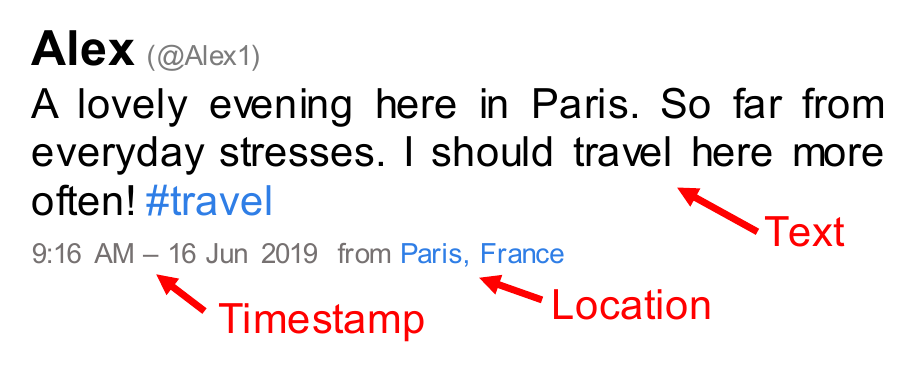}
	\captionsetup{font=footnotesize}
	\caption{Geo-tagged tweet: An example of a STT object}
	\label{fig:tweet}
\end{figure}

Due to the increased usage of mobile devices and advancements in accurate geo-tagging, more and more geo-tagged data is being produced~\cite{GeoTextualChallenges}.
In particular, social media platforms like Twitter and Facebook are some of the main sources of geo-tagged data, usually in the form of posts, comments, and reviews (e.g., Figure~\ref{fig:tweet}).
This type of data contains spatial, textual, and temporal (STT) information.
As a result, \emph{STT data} analysis is becoming increasingly important~\cite{QueryingGeoTextual} since it allows to extract new insights regarding customer satisfaction, user-generated content shared online, and brand reputation~\cite{userOpinions}.

\emph{STT data} contains information regarding topics discussed w.r.t. time and location, hence presenting an invaluable link between user opinions and the real world.
For example, STT data can help us analyze an advertisement campaign to identify the best locations for ad placements.
Traditionally, this information is accessed through spatial keyword-queries~\cite{SpatialQuery}, e.g., to retrieve topics within a certain location, or identify in which locations some topic is discussed.
However, keyword or topic search are \emph{point-wise} search tasks. Instead, there a significant need to \emph{provide more extensive analytics analogous to traditional OLAP-style analytics}.
An example STT query is \emph{``find the top-k trending hashtags aggregated by topic within a user-defined region (i.e., polygon) around Paris this month"}.

The traditional data cube model is one of the most widely used tools to analyze structured data. 
Since their introduction, data cubes have been extended to analyze different types of data, like sales~\cite{sweb}, locations~\cite{GeoMiner}, time-series~\cite{TimeSeries}, and text~\cite{TextCube}, but \emph{separately}.
In particular, some works propose OLAP operators to analyze either textual data~\cite{MDeXploration,newTextOLAP,MMMTD} or spatial data~\cite{sweb,GeoMiner}. 
However, no previous work proposed a unified model and set of operators enabling \emph{integrated} and \emph{joint} analysis of \emph{STT data}.
Moreover, as we propose to \emph{jointly} analyze STT dimensions together with other dimensions, we are also able to define novel families of measures that have not been studied before, namely \emph{spatio-textual} and \emph{spatio-textual-temporal measures}.
These measures, as we show later, allow to produce more advanced analytics instead of, e.g., simple keyword frequency.
\setlength{\textfloatsep}{3pt plus 0pt minus 0pt}
\begin{table}[b!]
	\scriptsize
	\begin{center}
		\captionsetup{font=footnotesize, skip=0pt}
		\caption{Spatio-Textual-Temporal Sample Dataset}
		\label{tab:data-sample}
			\begin{tabular}{|c|c|c|c|}
			\hline
			 & \textbf{Time} & \textbf{Location}  & \textbf{Terms} \\ \hline
			\hline
			1 & 11:12:13 20-10-2019 & 57.016254, 09.991203 & Apple, fruit, \#love \\ \hline
			2 & 11:18:23 24-10-2019 & 56.187421, 10.171410 & Potato, \#NewYear \\ \hline
			3 & 11:35:56 20-10-2019 & 56.151078, 10.204762 & Banana, Season \\ \hline
			4 & 16:12:14 24-10-2019 & 57.016254, 09.991203 & Potato, Salad, \#Fresh \\ \hline
			\dots & \dots & \dots & \dots
		\end{tabular}
	\end{center}
\end{table}
\definecolor{myRed}{RGB}{228, 26, 28}
\definecolor{myGreen}{RGB}{77, 175, 74}
\definecolor{myBlue}{RGB}{55, 126, 184}
\newcommand{\cmark}{\textcolor{myGreen}{\ding{52}}}
\newcommand{\xmark}{\textcolor{myRed}{\ding{56}}}
\newcommand{\win}{\cellcolor{myGreen!30}}
\newcolumntype{S}[1]{>{\centering\let\newline\\\arraybackslash\hspace{0pt}}m{#1}}
\begin{table*}[!t]
	\setlength{\belowrulesep}{0pt}
	\setlength{\aboverulesep}{0pt}
	\captionsetup{font=small, skip=0pt}
	\caption{Presence (\cmark) or absence (\xmark) of support for spatial and textual data, dimensions, hierarchies, and measures in existing methods}
	\label{tbl:relatedwork}
	\begin{center}
		{
			\tiny
			\newcolumntype{C}{>{\centering\arraybackslash}X}
			\begin{tabularx}{\textwidth}{p{3.5cm}CCCC|CCCC|C|C}
								\midrule
				&\multicolumn{4}{c|}{\textbf{Textual Support}}&\multicolumn{4}{c|}{\textbf{Spatial Support}}&\textbf{ST}&\textbf{STT}\\
				\textbf{Method} & \scriptsize{Data} & \scriptsize{Dimension} & \scriptsize{Hierarchies} & \scriptsize{Measures} & \scriptsize{Data} & \scriptsize{Dimension} & \scriptsize{Hierarchies} & \scriptsize{Measures} & \scriptsize{Measures} & \scriptsize{Measures}\\
				\midrule
				EXODuS~\cite{EXODuS} & \cmark (JSON) & \xmark &  \xmark & \xmark & \xmark & \xmark & \xmark & \xmark & \xmark & \xmark\\
				TextCube~\cite{TextCube} & \cmark & \cmark & \cmark & \cmark & \xmark & \xmark & \xmark & \xmark & \xmark & \xmark \\
				Text OLAP~\cite{TopicHierarchyModeling} & \cmark & \cmark & \cmark & \xmark & \xmark & \xmark & \xmark & \xmark & \xmark & \xmark \\
				TextCubeTopKCells~\cite{TextCubeTopKCells} & \cmark & \cmark & \cmark & \cmark & \xmark & \xmark & \xmark & \xmark & \xmark & \xmark \\
				Geo Miner~\cite{GeoMiner} & \xmark & \xmark & \xmark & \xmark & \cmark & \cmark & \cmark & \cmark & \xmark & \xmark \\
				SpatialCube~\cite{sweb} & \xmark & \xmark & \xmark & \xmark & \cmark & \cmark & \cmark & \cmark & \xmark & \xmark \\
				StreamCube~\cite{StreamCube} & \cmark & \xmark & \xmark & \cmark & \cmark & \cmark & \cmark & \xmark & \xmark & \xmark \\
				TwitterSand~\cite{TwitterSand} & \cmark & \cmark & \cmark & \xmark & \cmark & \xmark & \xmark & \cmark & \xmark & \xmark \\
				TextStreams~\cite{textStreams} & \cmark & \xmark & \xmark & \cmark & \cmark & \xmark & \xmark & \xmark & \xmark & \xmark \\
				TopicExploration~\cite{TopicExploration} & \cmark & \xmark & \xmark & \cmark & \cmark & \cmark & \cmark & \xmark & \xmark & \xmark \\
				SocialCube~\cite{TextCubeApproach} & \cmark & \xmark & \xmark & \cmark & \cmark & \cmark & \cmark & \xmark & \xmark & \xmark \\
				TopicCube~\cite{TopicCube} & \cmark & \cmark & \cmark & \cmark & \cmark & \cmark & \cmark & \xmark & \xmark & \xmark \\
				ContextualizedWarehouse~\cite{ContextualizedWarehouse} & \cmark & \xmark & \xmark & \cmark & \cmark & \cmark & \cmark & \xmark & \xmark & \xmark \\
				\midrule
				\rowcolor{myBlue!10}
				STTCube & \cmark & \cmark & \cmark & \cmark & \cmark & \cmark & \cmark & \cmark & \cmark & \cmark \\
				\bottomrule
			\end{tabularx}
		}
	\end{center}
\end{table*}

\para{Contributions.}
In this paper, we introduce the Spatio-Textual-Temporal Cube (STTCube) to analyze \emph{STT data}.
Adding spatial, textual, and temporal support to a traditional data cube is not straight-forward due to the presence of $n$$-$$n$ relationships in textual hierarchies and because existing families of measures cannot support \emph{joint} and \emph{integrated} analysis involving spatial, textual, and temporal dimensions, e.g., finding the trending keywords grouped by regions, defined by geometry shapes, over a time interval (Section~\ref{sec:STTMeasures}).
Hence, we introduce new families of measures and OLAP operators that extract \emph{combined insights} from STT dimensions and measures. 
STTCube provides specialized \emph{spatio-textual} and \emph{spatio-textual-temporal measures} such as \emph{Top-k Dense Keywords within an area} and \emph{Top-k Volatile Keywords within an area} that deliver the integrated aggregates over \emph{STT data}.
Moreover, a set of analytical operators, namely STT slice, dice, roll-up, and drill-down are proposed.
This results in a data model able to support \emph{spatio-textual-temporal OLAP} (STTOLAP) operators.
Furthermore, we propose \emph{Partial Exact Materialization (PEM)} and \emph{Partial Approximate Materialization (PAM)} methods for efficient exact and approximate computations of \emph{STT measures}, respectively.
Among other things, we also provide a systematic set of solutions to handle $n$$-$$n$ relationships in textual hierarchies.
In this work, we present the following contributions:
\begin{itemize}[leftmargin=*]
    \item We extend the standard cube model to add support for \emph{spatial, textual, and temporal} dimensions and hierarchies and \emph{spatio-textual} and \emph{spatio-textual-temporal} measures (\Cref{sec:STTCubeSchema,sec:SThierarchies,sec:STTMeasures}).
    \item We propose a set of analytical operators (STTOLAP) over \emph{spatio-textual-temporal data} (\Cref{sec:stolap}).
    \item We introduce \emph{keyword density} and \emph{keywords volatility} as prototypical \emph{spatio-textual} and \emph{spatio-textual-temporal measures} (\Cref{sec:STTMeasures}).
    \item We propose a pre-aggregation framework (STTCube materialization) for efficient, exact (\emph{(PEM)}) and approximate (\emph{(PAM)}), computation of the proposed \emph{spatio-textual-temporal measures} (\Cref{sec:greedy}).
    \item We propose techniques for processing \emph{spatio-textual-temporal objects} and the construction of the \emph{STTCube} (\Cref{sec:design}).
    \item We evaluate the pre-aggregation framework\textquotesingle s (\textit{PEM} and \textit{PAM}) query response time, storage cost, and accuracy by comparing it with the \textit{No STT Cube}, \textit{Full Materialization}, and \textit{No Materialization} baselines.
    Our pre-aggregation framework provides 1-5 orders of magnitude improvement in query response time and a 97\% to 99.9\% reduction in storage cost with an accuracy between 90\% and 100\% (\Cref{sec:experiments}).
\end{itemize}
\section{Related Work}\label{sec:related}
OLAP and the \emph{Data Cube}~\cite{dwbook} are used heavily in business intelligence to obtain insights over the historical, current, and future state of business.
With the emergence of web and social media, an immense amount of unstructured data is being produced, which must be included in the analytical process.
Table~\ref{tbl:relatedwork} summarizes the state of the art on spatial, textual, and temporal analytics by listing the properties and gaps in the current methods.

The \emph{Text-Cube}~\cite{TextCube} allows OLAP-like queries on text data by providing dimensions and hierarchies for terms.
Moreover, it supports the computation of two information retrieval (IR) measures: \emph{inverted index} and \emph{term frequency}. 
\emph{EXODuS}~\cite{EXODuS} processes semi-structured document stores (i.e., JSON) using a schema-on-read approach to allow exploratory OLAP on text.
Text OLAP~\cite{TopicHierarchyModeling} extends traditional OLAP to support textual dimensions and  keyword-based top-\emph{k} search~\cite{TextCubeTopKCells}.
\emph{Yet, all these approaches lack support for spatial and temporal data and the advanced measures and operators required for spatio-textual-temporal analytics.}

For spatial data, GeoMiner~\cite{GeoMiner} proposes a cube structure for mining characteristics, comparisons, and association rules from geo-spatial data and Spatial cube~\cite{sweb} allows to perform spatial OLAP on the semantic web.
\emph{Yet, these solutions focus on spatial data only and lack support for textual and temporal data.}

There are solutions that combine more than one component of data, e.g., spatio-temporal~\cite{spatioTemporal}, into the same model but do not provide combined \emph{STT} analytics.
Among those, the contextualized warehouse~\cite{ContextualizedWarehouse} combines traditional OLAP with a textual warehouse.
This allows the user to provide some keywords, select a market (country or region), retrieve documents matching the keywords as context, and then analyze the facts related to those keywords and documents.
Similarly, \emph{Topic Cube}~\cite{TopicCube} extends the functionality of a traditional cube and combines probabilistic topic modeling with OLAP by introducing the \emph{topic hierarchy}.
TwitterSand~\cite{TwitterSand} and StreamCube~\cite{StreamCube} exploit textual and spatial information to gain insights by clustering twitter hashtags and tweets in a region, respectively.
\emph{STT data} is also analyzed to extract events and topics information in TextStreams~\cite{textStreams} and TopicExploration~\cite{TopicExploration}.
Finally, SocialCube~\cite{TextCubeApproach} tries to capture human, social, and cultural behavior by performing linguistic analysis (sentiment analysis) over tweets.
All these approaches focus on the unstructured nature of text along with spatial and temporal data but they \emph{do not provide Integrated STT analytics}, for example, they do not provide the ability to \emph{compute aggregate spatial, textual, temporal, and spatio-textual-temporal measures over spatial, textual, and temporal dimensions and hierarchies}.

Spatial top-\textit{k} keyword-queries~\cite{QueryingGeoTextual,ExperimentalEvaluation,microblogsSurvey} \emph{answer only point-wise} queries and do not support aggregation functions or hierarchies.
Thus, they do not support more complex OLAP-style analytical tasks, which we do.
There are methods that solve a very specific task for a specific type of data~\cite{GeoTrend+,STEWARD,Waldo}. 
These methods are fundamentally different from STTCube because \emph{STTCube provides a generic framework for a wide range of STT analytics over different kinds of STT data sources}, including, but not limited to, geo-tagged tweets.
Also, STTCube can take advantage of the improvements suggested over other cubes, e.g., Nanocubes~\cite{Nanocubes} and DICE~\cite{DICE}, making it a powerful tool for OLAP-style \emph{STT analytics}.

Our summary of related work in Table~\ref{tbl:relatedwork} shows that no existing method provides integrated support for \emph{STT data}, unlike STTCube.
To the best of our knowledge, a proper formalization of a data cube model for \emph{STT data} able to support complex analytics for \emph{STT objects} at scale is missing.
In particular, no previous method studies dimensions, hierarchies, and measures that allow processing STT data \emph{jointly}. 
Furthermore, the main novel challenge for \emph{STT-OLAP} is handling $n$$-$$n$ relationships inside the \emph{STT} dimensions effectively since $n$$-$$n$ relationships do not allow traditional pre-aggregation techniques to be used.
Moreover, arbitrary temporal ranges with multiple levels of granularity adds complexity to \emph{STT} measures computations.
As a remedy, we propose \emph{STTCube} which enables the \emph{joint} and \emph{integrated} analysis of \emph{STT objects} by introducing new sets of measures, \emph{spatio-textual} and \emph{spatio-textual-temporal measures}, to gain in-depth insights using STTOLAP operators.  

\section{Spatio-Textual-Temporal Cubes} \label{sec:STCube}
Here, we define the \emph{STTCube}, an extension of the traditional data cube to allow storage and analysis of \emph{STT objects}.
Data cubes are used to model and analyze multi-dimensional data.

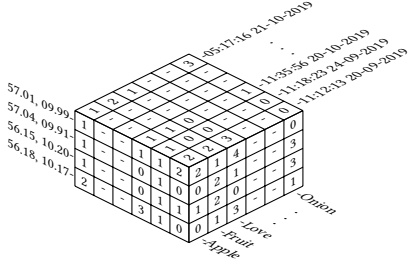
\begin {figure}[t!]
	\centering
		\begin{tikzpicture}[every node/.style={minimum size=1cm},on grid, scale=.25]
			\tikzstyle{every node}=[font=\tiny]
			\begin{scope}[every node/.append style={yslant=-0.5},yslant=-0.5]
				\node at (0.5,5.5) {1};
				\node at (1.5,5.5) {-};
				\node at (2.5,5.5) {-};
				\node at (3.5,5.5) {1};
				\node at (4.5,5.5) {1};
				\node at (5.5,5.5) {2};
				\node at (-1.8,5.5) {57.01, 09.99-};
				
				\node at (0.5,4.5) {1};
				\node at (1.5,4.5) {-};
				\node at (2.5,4.5) {-};
				\node at (3.5,4.5) {0};
				\node at (4.5,4.5) {1};
				\node at (5.5,4.5) {0};
				\node at (-1.8,4.5) {57.04, 09.91-};
				
				\node at (0.5,3.5) {1};
				\node at (1.5,3.5) {-};
				\node at (2.5,3.5) {-};
				\node at (3.5,3.5) {0};
				\node at (4.5,3.5) {1};
				\node at (5.5,3.5) {1}; 
				\node at (-1.8,3.5) {56.15, 10.20-};
				
				\node at (0.5,2.5) {2};
		  		\node at (1.5,2.5) {-};
				\node at (2.5,2.5) {-};
				\node at (3.5,2.5) {3};
		  		\node at (4.5,2.5) {1};
				\node at (5.5,2.5) {0};
				\node at (-1.8,2.5) {56.18, 10.17-};
				
				\draw (0,2) grid (6,6);
			\end{scope}
			\begin{scope}[every node/.append style={yslant=0.5},yslant=0.5] 
				\node at (6.5,-0.5) {2};
				\node at (7.5,-0.5) {1};
				\node at (8.5,-0.5) {4};
				\node at (9.5,-0.5) {-};
				\node at (10.5,-0.5) {-};
				\node at (11.5,-0.5) {0};
		
				\node at (6.5,-1.5) {0};
				\node at (7.5,-1.5) {2};
				\node at (8.5,-1.5) {1};
				\node at (9.5,-1.5) {-};
				\node at (10.5,-1.5) {-};
				\node at (11.5,-1.5) {3};
		
				\node at (6.5,-2.5) {1};
				\node at (7.5,-2.5) {2};
				\node at (8.5,-2.5) {0};
				\node at (9.5,-2.5) {-};
				\node at (10.5,-2.5) {-};
				\node at (11.5,-2.5) {3};
		
				\node at (6.5,-3.5) {0};
				\node at (7.5,-3.5) {1};
				\node at (8.5,-3.5) {3};
				\node at (9.5,-3.5) {-};
				\node at (10.5,-3.5) {-};
				\node at (11.5,-3.5) {1};
		
				\draw (6,-4) grid (12,0);
			\end{scope}
			\begin{scope}[every node/.append style={yslant=0.5,xslant=-1},yslant=0.5,xslant=-1] 
				\node at (6.5,5.5) {1};
				\node at (6.5,4.5) {-};
				\node at (6.5,3.5) {-};
				\node at (6.5,2.5) {1};
				\node at (6.5,1.5) {1};
				\node at (6.5,0.5) {2};
		
				\node at (7.5,5.5) {2};
				\node at (7.5,4.5) {-};
				\node at (7.5,3.5) {-};
				\node at (7.5,2.5) {1};
				\node at (7.5,1.5) {0};
				\node at (7.5,0.5) {2};
		
				\node at (8.5,5.5) {1};
				\node at (8.5,4.5) {-};
				\node at (8.5,3.5) {-};
				\node at (8.5,2.5) {0};
				\node at (8.5,1.5) {0};
				\node at (8.5,0.5) {3};
		
				\node at (9.5,5.5) {-};
				\node at (9.5,4.5) {-};
				\node at (9.5,3.5) {-};
				\node at (9.5,2.5) {-};
				\node at (9.5,1.5) {-};
				\node at (9.5,0.5) {-};
		
				\node at (10.5,5.5) {-};
				\node at (10.5,4.5) {-};
				\node at (10.5,3.5) {-};
				\node at (10.5,2.5) {-};
				\node at (10.5,1.5) {-};
				\node at (10.5,0.5) {-};
		
				\node at (11.5,5.5) {3};
				\node at (11.5,4.5) {-};
				\node at (11.5,3.5) {-};
				\node at (11.5,2.5) {1};
				\node at (11.5,1.5) {0};
				\node at (11.5,0.5) {0};
		
				\node[rotate=-90] at (2.5,-5.15){-Apple};
				\node[rotate=-90] at (3.5,-5.05){-Fruit};
				\node[rotate=-90] at (4.5,-5.05){-Love};
				\node[rotate=-90] at (6.25,-5.25){$\vdots$};
				\node[rotate=-90] at (7.5,-5.25){-Onion};
	
				\node at (15.05,0.5){-11:12:13 20-09-2019};
				\node at (15.05,1.5){-11:18:23 24-09-2019};
				\node at (15.05,2.5){-11:35:56 20-10-2019};
				\node at (14.75,4.25){$\vdots$};
				\node at (15.05,4.5){};
				\node at (15.05,5.5){-05:17:16 21-10-2019};
				
				\draw (6,0) grid (12,6);
			\end{scope}
		\end{tikzpicture}
	\captionsetup{font=footnotesize, skip=0pt}
	\caption{STTCube Example}
	\label{fig:st-cube-example}
\end{figure}

\textit{\textbf{Definition 1 (Data Cube).}}
\textit{An \textit{n-dimensional} data cube $\mathcal{CS}_{dc}$ is a tuple $\mathcal{CS}_{dc}$$=$$(D, M, F)$, with a set of dimensions $D$$=$$ \{d_1, d_2, \cdots,$ $d_n\}$, a set of measures $M$$=$$\{m_1, m_2, \cdots, m_k\}$, and a set of facts $F$.
A dimension $d_i$$\in$$D$ has a set of hierarchies $H_{d_i}$.
Each hierarchy $h$$\in$$H_{d_i}$ is organized into a set of levels $L_h$.
Each level $l$$\in$$L_h$ contains a set of members and has a set of attributes $A_l$.
Each attribute $a$$\in$$A_l$ is defined over a domain.
Each measure $m$$\in$$M$ is a function defined over a domain which can return either a single value or a complex object.
The domain of a dimension $d_i$ is denoted by $\delta\left(d_i\right)$ }

\para{Spatio-Textual-Temporal (STT) Objects}
An \emph{STT object} records place (geo-coordinates or location where it was created), text (a review, or a user comment), and time (when it was created).
Social networks with geo-tagged micro-blog posts are typical \emph{STT data} sources (e.g., the geo-tagged tweet in Figure~\ref{fig:tweet}).

\textit{\textbf{Definition 2 (STT object).}
A spatio-textual-temporal object is a tuple $obj_{st}$$=$${\langle}$$\lambda, \varphi, \tau$$\rangle$ where $\lambda$, $\varphi$, and $\tau$ represent the location, text, and time components, respectively.
}

The \textit{Location} is represented as the latitude and longitude pair $\lambda$$\in$$\left( \mathbb{R}\times\mathbb{R}\right)$.
The \textit{Text} is an ordered list $\varphi$$=$${\langle}w_1, w_2, \dots, w_n{\rangle}$ where $w_i$$\in$$\mathcal{W}$ is a string and is called a \emph{Term}.
Among all Terms, keywords are a user-defined subset of important Terms $W_k$$\subseteq$$\mathcal{W}$.
For instance, the user can decide that hashtags  (terms starting with $\#$) have special meaning and are a special type of keyword.
\textit{Time} specifies a precise instant (a timestamp) to some resolution (e.g., seconds).
Table~\ref{tab:data-sample} contains examples of \emph{STT objects} with their location, a set of keywords extracted from the text, and timestamp.

\subsection{The STTCube Schema} \label{sec:STTCubeSchema}
For analytical processing of \emph{STT objects} we propose to model them as an \emph{STTCube}. 
An STTCube~$\mathcal{CS}_{sttc}$$=$$($$D, M, F$$)$ is a data cube (Definition~1) with three special dimensions, namely \textit{Location}, \textit{Text}, and \textit{Time} that is $D = \{d_{Location}, d_{Text}, d_{Time}, d_{4}, \dots, d_{n}\}$.
\para{Dimensions.}
An \emph{STTCube} stores \emph{STT objects} as facts modeling their spatial, textual, and temporal features in the corresponding dimensions.
Figure \ref{fig:st-cube-example} shows a \textit{3-}dimensional \emph{STTCube} built on the sample dataset in Table~\ref{tab:data-sample} where each row represents one fact (i.e., the members of $F$) with dimensions $D$$=$$\{d_\textit{Location}, d_\textit{Text}, d_\textit{Time}\}$.
Domains for the respective dimensions are
\begin{align}
	\delta\left(d_\textit{Location}\right) &= \left\lbrace \textit{(57.016, 09.991)}, \textit{(56.187, 10.171)}, \dots\right\rbrace \nonumber\\
	\delta\left(d_\textit{Text}\right) &= \left\lbrace\textit{apple}, \textit{Fruit}, \textit{\#love}, \dots \right\rbrace \nonumber\\
	\delta\left(d_\textit{Time}\right) &=  \left\lbrace \textit{11:12:13 20-10-2019}, \textit{11:18:23 24-10-2019}, \dots\right\rbrace \nonumber
\end{align}
Hence w.r.t. Definition~2, the dimensions capturing $\lambda$, $\varphi$, and $\tau$ are the spatial, textual, and temporal dimensions, respectively.

\para{Dimension Hierarchy.}
A hierarchy is spatial, textual, or temporal if it contains spatial, textual, or temporal levels, respectively.
In Figure \ref{fig:st-cube-example}, the \emph{Location} dimension is a spatial dimension with a spatial hierarchy going from $\lambda$ to \emph{City, Region}, and \emph{Country} and the \emph{Text} dimension is a textual dimension aggregating $\varphi$ into the \emph{Term, Theme, Topic}, and \emph{Concept} levels.
Similarly, \emph{Time} is a temporal dimension.
Hierarchy steps $HS_h$$=$$\{hs_1, hs_2, hs_3, \dots, hs_n\}$ define the mechanism of moving from a lower (child) level to an upper (parent) level and vice versa.
A hierarchy step $hs_i$$=$$(l_\downarrow, l_\uparrow, car-$ $dinality)$$\in$$HS_h$ entails that members of a child level $l_\downarrow$ can be aggregated together if they correspond to the same member at the parent level $l_\uparrow$ and that this correspondence between children to parent members has the given $cardinality$$\in$$\{1$$-$$1, 1$$-$$n, n$$-$$1, n$$-$$n\}$.
For instance, the step from \emph{Date} to \emph{Month} has an $n$$-$$1$ cardinality, while \emph{Term} to \emph{Topic} has an $n$$-$$n$ cardinality (e.g., the Carrot \emph{Term} correspond both to the Gardening and Food \emph{Topics}, while the Food \emph{Topic} has as child members not only Carrot but also Apple).

\para{Level Attributes.}
As mentioned earlier, a level $l$ is associated with a set of attributes $A_l$$=$$\{a_1, a_2, \dots, a_n\}$ and has a set of members $l$$=$$\{l_1, l_2, \dots, l_3\}$.
Attribute values describe the different characteristics of each member from that level.
Spatial, textual, and temporal levels are then usually characterized by spatial, textual, and temporal attributes.
For instance, at the \emph{City} level, member Aalborg has the \emph{Boundary} attribute whose value is the polygon defining the boundary of Aalborg. 
An example of a textual attribute is \emph{Sentiment} which captures the polarity of the associated textual member.
Similarly, an \emph{integer value} representing the number of days in a specific month is a temporal attribute.

\subsection{Managing STT Hierarchies}\label{sec:SThierarchies}
We now describe the STTCube\textquotesingle s dimensions and hierarchies.

\para{Spatial Dimensions.}
Spatial information can be analyzed at different levels and granularities.
It is important to note that facts in an STTCube are composed only by geographical \emph{points} (i.e., each tweet or user post is associated with a coordinate, not with shapes or polygons).
Points can be aggregated either within a predefined spatial grid or based on semantic information.

\textit{Grid-Based Hierarchy.}
Here, the geographic area being analyzed is divided into small equal size cells with a predefined resolution, e.g., $1$$\times$$1\,km^{2}$.
At the lowest level, each latitude and longitude point is assigned to the cell they fall in.
To analyze data at a coarser granularity, neighboring cells are combined into a larger cell at the parent level (e.g., $3$$\times$$3\,km^2$).
This hierarchy can be built automatically, without the need for any meta-data.

\textit{Semantic-Based Hierarchy.}
Here, data is analyzed in a predefined taxonomy, e.g., an administrative division.
Therefore, we move within the taxonomy, e.g., from the \textit{Location} to the \textit{City} level, from the \textit{City} level to the \textit{Region} level, and so on up to the \textit{All} level.
This hierarchy requires each object coordinate to be associated with a member in the lowest level in the hierarchy (usually in a pre-processing step) and requires the taxonomy information to build the entire hierarchy.

\para{Textual Dimensions.}
Hierarchies in the textual dimension move from specific concepts to general ones.
This follows a generic taxonomic structure connecting more specific terms to more general ones (i.e., hypernyms)~\cite{taxonomy}.
In particular, \textit{Terms} are the base level which are grouped into \textit{Themes}, \textit{Themes} into larger categories called \textit{Topics}, and \textit{Topics} in turn grouped into \textit{Concepts}.
\emph{Differently from most hierarchies, the members in the levels of a textual hierarchy are typically in an $n$$-$$n$ relationship.}
Hence, when moving between textual levels we need to decide how measure values get aggregated. 
Below we propose a set of aggregation techniques to address this issue.

\textit{Replication-Based Hierarchy.}
This is a common approach where each member of a child level is aggregated into all the parent members. Hence, its value is effectively replicated.
This approach leads to a \emph{counting problem} when parent levels are further aggregated.
For example, the first data instance in Table~\ref{tab:data-sample} will be part of two \textit{Themes}: 1) Fruits because it contains \textit{Term} $\{$apple and fruit$\}$ and 2) Emotion because of \textit{Term} $\{$\#love$\}$.

\textit{Majority-Based Hierarchy.}
If a fact can be mapped to more than one parent member, then that fact will be part of the parent member which has the most representation (e.g., in terms of frequency).
This scheme avoids double counting of facts in parent members.
In case of ties, some tie-breaking heuristic or a user-defined criterion can be employed instead, e.g., the first fact in Table~\ref{tab:data-sample} will be part of only the Fruits \textit{Theme}  because it has the two representative \emph{Term} $\{$apple, fruit$\}$, as compared to Emotion having only one \textit{Term} $\{$\#love$\}$.

\textit{Custom Hierarchy.}
In general, other user-specified criteria and rules can be defined to establish how child-parent level steps will be aggregated in case of ambiguities.
For instance, a domain-specific \emph{importance score} can be assigned to the hierarchy members during the \emph{STTCube} construction. 
In this way, facts will be part of only the parent member with the highest importance.

\para{Temporal Dimensions.}
Similarly, temporal dimensions allow to analyze \emph{STT objects} at different levels of granularity w.r.t. time and has the following two temporal hierarchies: $\tau \rightarrow Day \rightarrow Month \rightarrow Quarter \rightarrow Year \rightarrow all$ and $\tau \rightarrow Second \rightarrow Minute \rightarrow Hour \rightarrow all$.
{\noindent}Here, the first contains a hierarchy of \emph{Date} aggregated by the temporal levels \emph{Day, Month, Quarter}, and \emph{Year} (total 5 levels including \textit{All}), whereas the second is a hierarchy for \emph{TimeOfDay} having 4 levels in total.

\subsection{Spatial, Textual, and Temporal Measures}\label{sec:STTMeasures}

As defined earlier, an \textit{n-dimensional} \emph{STTCube} has a set of measures $M$$=$$\{m_{1}, m_{2}, m_{3}, \dots ,m_{k}$$\}$, which permit to analyze \emph{STT objects} by computing values at different levels of granularity.
For instance, the \emph{STTCube} in Figure \ref{fig:st-cube-example} models \textit{Location}, \textit{Text}, and \textit{Time} with \emph{Fact Count} as a measure (i.e., {\textit{Fact Count}}$\in$$M$).
In practice, it maintains the count of \emph{STT objects} at given spatial, textual, and temporal aggregation levels.
Measure values at different levels in the hierarchies are obtained by applying an aggregation function over the \emph{STT objects}.
Examples of aggregation functions are \emph{SUM}, \emph{COUNT}, \emph{MIN}, \emph{MAX}, and \emph{AVG}.
The \emph{STTCube} in Figure \ref{fig:st-cube-example} uses \textit{COUNT} as an aggregation function.
For example, it reports that on \emph{September, 20th} at \textit{AAU Bus Terminal} the \textit{Term} \textit{apple} was mentioned in 2 facts.

A measure is spatial if it is defined over a spatial domain.
A spatial measure is then computed over a collection of spatial values (e.g., geographical points, or geometry shapes like polygons).
A spatial measure can be a simple value, e.g., the (numeric) area of the convex hull of multiple shapes, or a complex spatial object, e.g., the polygon representing the convex hull itself.
A measure is textual if it is defined over a textual domain, and can be either a simple numeric value or a complex textual object.
Analogously, a measure is temporal if it is defined over a temporal domain, 
A measure is spatio-textual if it is defined over a spatial and textual domain and is a combination of spatial and textual measures.
Finally, a measure is spatio-textual-temporal if it is defined over a spatial, textual, and temporal domain and is a combination of spatial, textual, and temporal measures.
Below, we propose a list of \emph{spatio-textual and spatio-textual-temporal measures} to be used as part of \emph{STTCube} to analyze \emph{STT objects} effectively.


\noindent\textbf{Top-k Keywords within an Area}
is a spatio-textual measure which returns a list of tuples ${\langle}\xi,\overrightarrow{kw}{\rangle}$ consisting of a geometry shape $\xi$ representing a geographical area and the list of top-\textit{k} most frequent keywords $\overrightarrow{kw}$$=$${\langle}w_1,w_2,\dots,w_k{\rangle}$ in that area.
Analogous to previous measures, It can also be computed at different levels of aggregation, so that it can return the top-k keywords for each \textit{City} or each \textit{Region}.

\noindent\textbf{Keyword Density}
is a spatio-textual measure which returns a list of tuples ${\langle}\xi_i, w_j, \rho_{ij}{\rangle}$ consisting of a geometry shape $\xi_i$ representing a geographical area, a keyword $w_j$, and its density $\rho_{ij}$ in the area $\xi_i$.
The density $\rho_{ij}$ of a keyword $w_j$ over an area $\xi_i$ is
computed as $\rho_{ij}=\frac{\textit{freq}({\xi_i,w_j})}{\textit{SurfaceArea}(\xi_i)}$, in which \textit{freq}$({\xi_i, w_j})$ is the frequency of the keyword $w_j$ in the area $\xi_i$ (i.e., the number of objects located within $\xi_i$ in which $w_j$ appears) and \textit{SurfaceArea} is the surface area of $\xi_i$. 
For example, if we have two \textit{Regions} $r_1$, $r_2$ with \textit{SurfaceArea}$(r_1)$$=$$10$$m^2$, \textit{SurfaceArea}$(r_2)$$=$$100$$m^2$, and the term \textit{Apple} with frequency $5$ and $30$ in $r_1$ and $r_2$, respectively (see Figure~\ref{fig:ATK}),
then, keyword densities are $\rho_1$$=$$0.5$, $\rho_2$$=$$0.3$ for $r_1$ and $r_2$, respectively.

\noindent\textbf{Top-k Dense Keywords within an Area}
is a spatio-textual measure which returns a list of tuples ${\langle}\xi_i, \overrightarrow{kw}{\rangle}$ computing the keyword density as described in the measure above, but in this case, it returns the top-k keywords $\overrightarrow{kw}$$=$${\langle}w_1,w_2,\dots,w_k{\rangle}$ with the highest density.

\noindent\textbf{Keyword Volatility}
is a spatio-textual-temporal measure (becomes textual-temporal If no region is specified) which returns a list of tuples ${\langle}\xi_i, w_j, T_k, \Delta\rho_{ijk}{\rangle}$ consisting of a geometry shape $\xi_i$ representing a geographical area, a keyword $w_j$, a time interval $T_k$, and its change in density $\Delta\rho_{ijk}$ in the area $\xi_i$ over the time interval $T_k$ (divided into $k$ equal intervals).
The change in density $\Delta\rho_{ijk}$ of a keyword $w_j$ in an area $\xi_i$ over a time interval $T_k$ is
computed as $\Delta\rho_{ijk}$$=$$\frac{\sum_{z=1}^{k}|\rho_{ij_{_z}} - \rho_{ij_{_{z-1}}}|}{k}$, where $\rho_{ij_{_z}}$ represents the density of the keyword $w_j$ in the area $\xi_i$ at a specific time instance $T_{k_z}$.
Furthermore, the change in density computation formula can be updated depending on the analysis requirements, e.g., it can be changed to weighted density (assign different weights to each interval in $T_k$) or to rate of change computation using linear regression~\cite{RateOfChange}.

\noindent\textbf{Top-k Volatile Keywords within an Area}
is a spatio-textual-temporal measure which returns a list of tuples ${\langle}\xi_i, \overrightarrow{kw}{\rangle}$ computing the keyword volatility as described above, but in this case, it returns the top-k volatile keywords $\overrightarrow{kw}$$=$${\langle}w_1,w_2,\dots,w_k{\rangle}$ with the highest change in density.

\noindent\textbf{Distributive, Algebraic, and Holistic Measures.}
There are three types (also known as additivity) of measures: distributive, algebraic, and holistic, depending on whether it is possible to compute the value of a measure at a parent level directly from the values at the child level~\cite{JimGray}.
For distributive and algebraic measures, this is possible.
For instance, the \textit{Fact Count} at the \textit{State} level can be computed by summing up the \textit{Fact Counts} at the \textit{City}. 
\textit{Keyword Density} is instead an algebraic measure.
We can compute the higher-level aggregate values of this measure if we store for each child level both the frequency of each keyword and the \textit{SurfaceArea}.
The \textit{Top-k Keywords}, the \textit{Top-k Dense Keywords}, and \textit{Top-k Volatile keywords within an area} measures, instead, are holistic, since the value at a parent level cannot be computed directly from the values at the child level, but it is necessary to recompute them directly from the base facts every time. 

Consider the computation of \textit{Top-3 Dense Keywords within an Area} in Figure~\ref{fig:ATK} given the two \textit{Regions} $r_1$ and $r_2$ with \textit{SurfaceArea} $10 m^2$ and $100 m^2$, respectively, and the computation at the parent level $r_3$$=$$r_1$$\cup$$r_2$ (grayed-out rows are not part of the computed measure value).
The values in the top-3 for the members $r_1$ and $r_2$ at the child level are not sufficient to compute the correct densities for region $r_3$.
Both, some of the computed density (in column $\rho_{Top-3}$, while the correct values are reported in $\rho_{all}$) and consequently the final ranking, would be wrong.
For instance, the keyword \textit{Strawberry} would not have been returned (if computed algebraically) because it is neither in the top-3 for $r_1$ nor $r_2$.
To compute the correct response, either we have to store all the aggregate values for each possible cell or we have to reprocess all the facts covered by the query.
When dealing with large datasets these approaches are not feasible.
Hence, in Section~\ref{sec:greedy} we provide a framework for the computation of an exact and approximate solution with accuracy guarantees.
\definecolor{Gray}{gray}{0.9}
\begin{figure}[t!]
	\centering
	\tiny
	\setlength{\tabcolsep}{1pt}
	\begin{tikzpicture}[sibling distance=50mm,edge from parent/.style={draw,latex-}]
	\node {
		\begin{tabular}{@{}cccccc@{}}
			\multicolumn{6}{@{}c}{Region $r_3= r_1 \cup r_2$}\\
			\toprule
			Keyword & $\Sigma_{_{all}}$ & $\Sigma_{_{Top-3}}$ & Area & $\rho_{_{all}}$ & \textcolor{red}{$\rho_{_{Top-3}}$}\\
			\midrule
			Carrot & 42 & 40 & \multirow{3}{*}{\rotatebox[origin=c]{270}{$110\,\,m^2$}} & 0.38  & \textcolor{red}{\underline{0.36}}\\
			Apple & 35 & 35 &  & 0.32 &  0.32\\
			Strawberry & 22 & 00 &  & 0.20 & \textcolor{red}{\underline{0.00}} \\
			\rowcolor{Gray}
			Banana & 20 & 20 &  & 0.18 &0.18\\
			\rowcolor{Gray}
			Orange & 16 &  05 & & 0.15 & \textcolor{red}{\underline{0.05}}\\
			\rowcolor{Gray}
			Potato & 04 & 04 & & 0.04& 0.04\\
			\bottomrule
		\end{tabular}
	}
	child {node[left=1.15cm, below=0.2cm] {
			\begin{tabular}{@{}cccc@{}}
				\multicolumn{4}{@{}c}{Region $r_1$}\\
				\toprule
				Keyword & Count & Area & Density\\
				\midrule
				Apple & 5 & \multirow{4}{*}{\rotatebox[origin=c]{270}{$10\,\,m^2\,\,$}}  & 0.50\\
				Orange & 5 &  & 0.50\\
				Potato & 4 &  & 0.40\\
				\rowcolor{Gray}
				Strawberry  & 3 & & 0.30\\
				\rowcolor{Gray}
				Carrot & 2 & & 0.20\\
				\bottomrule
			\end{tabular}
		}
	}
	child {node[right=1.15cm,below=0.2cm] {
			\begin{tabular}{@{}cccc@{}}
				\multicolumn{3}{@{}c}{Region $r_2$}\\
				\toprule
				Keyword & Count & Area & Density\\
				\midrule
				Carrot & 40 &  \multirow{4}{*}{\rotatebox[origin=c]{270}{$100\,\,m^2\quad$}} & 0.40\\
				Apple & 30 &  & 0.30\\
				Banana & 20 & & 0.20\\
				\rowcolor{Gray}
				Strawberry & 19 & & 0.19\\
				\rowcolor{Gray}
				Orange & 11 & & 0.10\\
				\bottomrule
				\end{tabular}
			}
		};
	\end{tikzpicture}%
	\captionsetup{font=footnotesize, skip=0pt}
	\caption{Example: Merging of Holistic Measure}
	\label{fig:ATK}
\end{figure}

\subsection{STTOLAP Operators} \label{sec:stolap}
A data cube allows different \textbf{O}n\textbf{L}ine \textbf{A}nalytical \textbf{P}rocessing (OLAP) operators to group, filter, and analyze cells and subsets of cells at different levels of granularity and under different perspectives. 
Those operators are known as \emph{Slice}, \emph{Dice}, \emph{Roll-Up}, and \emph{Drill-Down}~\cite{dwbook}. 
We extend the basic OLAP operators to \emph{STT-OLAP operators}, i.e., for spatial, textual, and temporal dimensions, hierarchies, and measures (Handing of $n$$-$$n$ relationships is explained in Section \ref{sec:greedy} and \ref{sec:design}).
In general, an OLAP (and STTOLAP) operator \textit{OP} accepts as input a cube $C'$, some parameters \textit{params} and outputs a new cube $C''$, i.e., \textit{OP}$(C'$, \textit{params}$)$$=$$C''$.
In this way, a new OLAP operator can be applied to $C''$.
Among all cubes, we distinguish the initial or \emph{base cube} $C$ as the cube containing all the original information at the base level.

\section{Cube Materialization}
\label{sec:greedy}
Cube materialization is the process of pre-aggregating measure values at different levels of granularity in the cube to compute query responses from pre-aggregated results instead of the raw data, and hence improve query response time for \emph{STTOLAP operators}~\cite{ImpCubeEff}.
In a data cube, a \emph{cuboid} is a collection of \emph{level members} and associated \emph{measure values} for a unique combination of dimension hierarchy levels. Each unique combination is represented by a separate cuboid. 
For instance, if we request the \emph{Fact Count} for the \emph{State} of Denmark and have stored \emph{Fact Count} at the \emph{Region} level, we can avoid accessing the raw data and compute the aggregation from much fewer rows.
This is an example of \emph{partial materialization}, i.e., the actual cuboid at the \emph{State} level, containing the answer to the query was not materialized, but the system was still able to exploit the cuboid for \emph{Region}.

What to materialize and how much to materialize depends on the trade-off between query response time and storage cost.
\textit{Full Materialization (FM)} is obtained by pre-computing measure values for \textit{all} combinations of levels in \textit{all} hierarchies.
This approach requires huge storage but achieves the best query response time since every operation can just look up already pre-computed results.
At the other extreme, \textit{No Materialization (NM)} only materializes the base cuboid and does not require any extra space, but will require aggregated measure values to be recomputed from the base cuboid every time, hence incurring much slower response times.
A middle-ground solution is to partially materialize the cube, i.e., to materialize only some of the possible cuboids.
In this strategy, some queries will be able to exploit pre-aggregated values at the current level, while other queries can exploit pre-aggregated values at lower levels for distributive or algebraic measures.



\subsection{Cost Model}
\label{sec:costmodel}
The core of the proposed \emph{partial materialization} approach depends on the trade-off between the storage cost of materializing any particular cuboid and the actual \emph{benefit} that the materialization of the cuboid provides.
To evaluate this benefit, we have to estimate the (run time) cost of a query.
To devise a cost model for this estimation, we performed a micro-benchmark which confirmed that the running time is directly proportional to the data size (the number of rows). Hence we can use the following linear cost model for benefit calculation\footnote{Due to space restrictions, the details and figures of the cost model experiments are available in \Cref{sec:latticeExampleApp,sec:costmodelApp}.}
$$\textit{Benefit}(c) = \sum_{c' \in \textit{descendants}(c) \cup \{c\}} \textit{cost}(c')-\textit{size}(c)$$


\subsection{Partial Exact Materialization}
We propose an exact partial materialization technique for pre-computing the \emph{spatio-textual} and \emph{spatio-textual-temporal measure} values.
To answer an \emph{STT query} for these measures we materialize two other distributive measures, namely \textit{Keyword Frequency} $f$ and \textit{SurfaceArea} $a$.
Then, since \textit{Keyword Density} $\rho$ and \textit{Keyword Volatility} $\Delta\rho$ are algebraic measures, they can be computed from the values of \textit{Keyword Frequency} $f$ and \textit{SurfaceArea} $a$.
Finally, \textit{Top-k Dense Keywords} and \textit{Top-k Volatile keywords} are holistic but for an exact solution we materialize \emph{Top-ALL} and hence, compute it from the materialized measure values (Figure~\ref{fig:ATK}).

\begin{algorithm}[b]
	\scriptsize
	\captionsetup{font=scriptsize, skip=0pt}
	\caption{Greedy  Materialization}
	\label{algo:GHM}
	{
		\SetAlgoLined
		\SetKw{Continue}{continue}
		\SetKwInput{KwInput}{Input}                
		\SetKwInput{KwOutput}{Output}              
		\SetKwProg{GreedyMaterialization}{GreedyMaterialization}{}{}
		\SetKwRepeat{Do}{do}{while}
		\GreedyMaterialization{$(B$, {\ccube}$, K)$}{
			\KwInput{Budget $B$, STTCube {\ccube}, desired top-k $K$}
			\KwOutput{Partially Materialized STTCube {\ccube}}
			\Do{\textit{size}$($\ccube$){\leq}B$}{
				\textit{Candidates}$\leftarrow\{$$V$$\in${\ccube}$| {\neg}V.$\textit{isMaterialized}$\}$\;
				$\overline{V}\leftarrow \max_{{V} \in \textit{Candidates}} \textit{Benefit}({V})$\;
				{\ccube}.\textit{materialize}$(\overline{V}, K)$\;
			}
			\Return{{\ccube}};
		}
	}
\end{algorithm}

We adopt the chosen linear cost model (Section~\ref{sec:costmodel}) and extend the greedy algorithm approach~\cite{ImpCubeEff} to our task (Algorithm~\ref{algo:GHM}).
\emph{Additionally, and different from~\cite{ImpCubeEff}, Algorithm~\ref{algo:GHM} accepts an input parameter $K$ and materializes only the top-\textit{K} measures values in each cuboid}.
For instance, for $K=10$, it will materialize the top-$10$ keywords in each cuboid. 
Then, any top-k query, with $k{\leq}K$, for a materialized cuboid will return the pre-computed answer. 

Algorithm~\ref{algo:GHM}, given a size budget $B$ (measured in rows, cuboids, or GB), proceeds until the size of the current cube is as large as possible within the budget (Line~6).
At each step, it selects among all the non-materialized cuboids (Line~3) the one with the highest benefit (Line~4) and  materializes it (Line~5).
The difference between the exact (PEM) and approximate (PAM) materialization using Algorithm~\ref{algo:GHM} is the value of $K$. 
When $K$$=$$\infty$ the full sorted list of measure values will be stored so that all top-\textit{k} queries can be answered for that cuboid.
We set $K$$=$$\infty$ and $K$$=$$n$ (to materialize only top $n$ measure values) for \emph{PEM} and \emph{PAM}, respectively.

\noindent\textbf{Query rewriting.}
Finally, as in~\cite{ImpCubeEff}, after \emph{STTCube} materialization, queries are still formulated in terms of the base cuboid but rewritten by the system to be evaluated over the smallest cuboid.
\begin{algorithm}[t!]
	\scriptsize
	\SetAlgoLined
	\SetKw{Continue}{continue}
	\SetKwInput{KwInput}{Input}                
	\SetKwInput{KwOutput}{Output}              
	\SetKwProg{TopKVolatile}{TopKVolatile}{}{}

	\TopKVolatile{$(\Phi$$=$$\{{\langle}\xi_1, \overrightarrow{kw_1},T_1{\rangle},\dots,{\langle}\xi_n, \overrightarrow{kw_n},T_n{\rangle} \}$,$T_x$,$k)$}{
		\KwInput{Set of Top-k+1 Volatile Keywords lists $\Phi$, Set of $x$ Timestamps $T_x$, Integer k}
		\KwOutput{${\langle}\xi, \overrightarrow{kw}, T_n{\rangle}$ top-k keywords $\overrightarrow{kw}$ in the merged area $\xi$ over time interval $T_x$, $\delta$ number of guaranteed top positions }
		
		$\xi \leftarrow \bigcup_{i{\in}[1,n]} \xi_i$, $A \leftarrow$\textit{SurfaceArea}$(\xi)$ \tcp*{Merge areas}
		$\overline{kw} \leftarrow \{\}$, $\Delta f \leftarrow \{\}$, $prev_{_f} \leftarrow \{\}$ \tcp*{Empty dictionaries}
		\ForEach{$t \in T_x$}{
			\ForEach{${\langle}\xi_i, \overrightarrow{kw_i}, T_i{\rangle} \in \Phi$}{
				\ForEach{$j \in [$$1,\dots,k$$+$$1$$]$}{
					\If{$t \in T_i$}
						{$w$$\leftarrow$$\overrightarrow{kw}_{i}.get(j)$\tcp*{keyword at $j$}
						$f$$\leftarrow$$\overrightarrow{kw}_i$.\textit{freq}$(j)$\tcp*{frequency at  $j$}
						$\overline{kw}[w] \leftarrow \overline{kw}[w]+f$\;
						$\Delta f[w] \leftarrow \Delta f[w] + |prev_{_f}[w] - f|$\;
						$prev_{_f}[w] \leftarrow f$\;
					}
				}
				$\epsilon \mathrel{+}= \overrightarrow{kw}_i$.\textit{freq}$(k+1)$\;
			}
		}
		$\overrightarrow{kw}\leftarrow$\textit{topK}$(\overline{kw},A,\Delta f)$ \tcp*{top-k volatile keywords}
		$\delta \leftarrow \max_{j \in [1,\dots,k]}{\overrightarrow{kw}.\textit{freq}(j) \geq \epsilon}$ \;
		
		\Return{${\langle}\xi, \overrightarrow{kw}, T_n{\rangle}$, $\delta$}
	}
	\captionsetup{font=scriptsize, skip=0pt}
	\caption{Top-K Volatile Keywords in an Area}
	\label{algo:STQP}
\end{algorithm}

\subsection{Partial Approximate Materialization}
\label{sec:approximate}
As a result of the materialization performed by Algorithm~\ref{algo:GHM}, when querying a non-materialized cuboid, we can directly exploit values in the cuboid's materialized ancestors when computing all distributive and algebraic measures.
On the other hand, for holistic measures, we have to perform some additional computation.
For instance, as mentioned earlier, to compute the value for the \textit{Top-k Dense Keywords in an area} we can exploit the pre-computed \textit{Keyword Density} values, but then we need to perform the top-\textit{k} selection.
That is, if the top-\textit{k} for the current view is not materialized, we cannot exploit the materialized top-\textit{k} of the ancestor views without incurring the risk of returning the wrong result.

Yet, it is possible to exploit the top-\textit{k} computation in some materialized cuboid to retrieve an \emph{approximate} top-\textit{k} and estimate the result\textquotesingle s accuracy~\cite{TopTerms}.
In practice, for the \textit{Top-k Dense Keywords within an area}, given a target \textit{k} for the top-\textit{k} computation, when materializing a cuboid, we materialize the top-\textit{k}+1 most dense keywords for that cuboid (i.e., set $K$=$k$+$1$ in Algorithm~\ref{algo:GHM}). 
Then, to compute the top-\textit{k} dense keywords for a descendant cuboid by exploiting a materialized ancestor cuboid, we determine which members of the list are guaranteed to be correct.

Algorithm~\ref{algo:STQP} implements this computation for \emph{Top-k Volatile Keywords within an area}.
It receives as input the set $\Phi$$=$$\{{\langle}\xi_1, \overrightarrow{kw_1},$ $T_1{\rangle},$$ {\langle}\xi_2, \overrightarrow{kw_2}, T_2{\rangle}, \dots, {\langle}\xi_n, \overrightarrow{kw_n}, T_n{\rangle} \}$ of lists of top-K (i.e., \textit{k}+1) dense keywords in a specific area with respective time stamps, time interval $T_x$ divided into x equal-sized interval (e.g., day or month), and the value for $k$.
The output is the ranked list of top-\textit{k} volatile keywords in the area $\xi$ that is composed by the merging of the areas $\xi_1,\xi_2,\dots,\xi_n$.
It computes the \textit{SurfaceArea} of the merged area $\xi$ (lines~2).
Then it merges all the aggregated keyword frequencies (line~10) and change in keyword frequencies (line~11) for each time instance in $T_x$(line~7) in respective dictionaries $\overline{kw}$ and $\Delta f$ (lines~4-13) by getting each keyword in each list (line~8) and the corresponding frequencies (line~9).
If a keyword is not found in the $\overline{kw}$, $\Delta f$, or $prev_{_f}$ dictionary then its value is considered to be \emph{zero}.
Moreover, it keeps track of the upper-bound $\epsilon$ frequency for keywords outside the current materialized ranking for possible error reporting (line~13).
Once all frequencies and changes in frequencies are merged, we compute the top-\textit{k} volatile keywords using the aggregated values (line~14). 
Finally, by comparing the value of $\epsilon$ with the frequencies of keywords in the aggregated top-\textit{k}, we report how many positions in the current ranking are guaranteed to be exact (line~15).
In the best case, the frequency of the keyword at position $k$ will be at least $\epsilon$ and thus the computed top-\textit{k} is guaranteed to be correct.


\section{STTCube Construction}
\label{sec:design}

Here, we describe the proposed approach for constructing an \emph{STTCube}. 
Algorithm \ref{algo:CC} takes a collection $\mathcal{X}$ of \emph{STT objects} to be analyzed, a textual taxonomy $\mathcal{T}$ with semantic information about the terms, themes, topics, and concepts, and a geographical taxonomy $\mathcal{G}$ for cities, regions, and countries.
Standard \emph{date functions} are used for the temporal dimension processing.
Moreover, it also receives as input the parameters $B$ and $K$ as the budget and number of top-\textit{K} keywords for the partial materialization.

\begin{algorithm}[!t]
	\scriptsize
	\SetAlgoLined
	\SetKwInput{KwInput}{Input}
	\SetKwInput{KwOutput}{Output}
	\SetKwProg{ConstructSTTCube}{ConstructSTTCube}{}{}
	
	\ConstructSTTCube{$(\mathcal{X},\mathcal{T},\mathcal{G}, B, K)$}{
		\KwInput{Collection of Spatio-Textual-Temporal Objects $\mathcal{X}$, Knowledge Source $\mathcal{T}$, Geographical Information $\mathcal{G}$, Materialization Budget $B$, desired top-k $K$}
		\KwOutput{Spatio-Textual-Temporal Cube {\ccube}}
		
		{\ccube}$\leftarrow $ load empty \textbf{or} existing cube\;
		{\ccube}$.d_\textit{Time} \leftarrow$ initialize \textbf{or} load temporal dimension\;
		{\ccube}$.d_\textit{Location} \leftarrow$ initialize \textbf{or} load spatial dimension\;
		{\ccube}$.d_\textit{Text} \leftarrow$ initialize \textbf{or} load textual dimension\;
		{\ccube}$.F \leftarrow$ initialize empty \textbf{or} load existing \textit{Fact Table}\;

		\ForEach{$x \in \mathcal{X}$}{
			\texttt{UpdateTemporalHierarchies}$(x.\tau,$ {\ccube}$.d_\textit{Time}$)\;
			$\lambda'\leftarrow$\texttt{ProcessLocation}$(x.\lambda)$\;
			\texttt{UpdateSpatialHierarchies}$(\lambda', \mathcal{G},$ {\ccube}$.d_\textit{Location}$)\;
			$\varphi'\leftarrow$\texttt{ProcessText}$(x.\varphi)$\;
			\texttt{UpdateTextualHierarchies}$(\varphi', \mathcal{T},$ {\ccube}$.d_\textit{Text}$)\;
			\texttt{InsertFact}($x.\tau, \lambda', \varphi'$, {\ccube}$.F$)\;
		}
		\texttt{GreedyMaterialization}$(B$, {\ccube}$, K)$\;
		\Return{{\ccube}}\;
	}
	\captionsetup{font=tiny}
	\caption{STTCubeConstruction}
	\label{algo:CC}
\end{algorithm}

Algorithm~\ref{algo:CC} constructs the STTCube in an incremental way, it initializes an empty cube (line~2), and then the corresponding spatial, textual, and temporal dimensions (lines~3-5) as well as the \textit{Fact Table} (line~6).
If the cube is already constructed, i.e., the cube is being updated instead of constructed for the first time, then Algorithm~\ref{algo:CC} loads the existing \emph{STTCube} (lines 2-6) and updates it with new information.
In particular, the spatial dimension has the grid-based hierarchy and the hierarchy with the base level at each object\textquotesingle s \textit{Location} (i.e., the geographical point), and then the levels \textit{City}, \textit{Region}, \textit{Country}, and \textit{All} (5 levels in total).
The textual dimension, instead, has the hierarchy build from the base level \textit{Term}, and then \textit{Theme}, \textit{Topic}, \textit{Concept}, and \textit{All} (5 levels in total).
Finally, the temporal dimension contains the \textit{Date} and \textit{TimeOfDay} hierarchies mentioned in Section~\ref{sec:STCube}.

Once the basic structure is prepared, Algorithm~\ref{algo:CC} loops through each STT object in $\mathcal{X}$ (lines~8-13).
In this loop, it extracts and initializes from each \emph{STT object} the base-level members for each dimension.
Then, once the base level data has been extracted, it proceeds with building the various dimension hierarchies starting from the existing base-level members and exploiting the provided spatial and textual taxonomies (lines~8-12).
Once the dimension hierarchies are built, the \emph{STT object} itself is then inserted in the fact table of the \emph{STTCube} (line~13) so that each fact is linked to the lowest (base) level members in the respective dimensions.
In this step (line~13), the fact measure values are also computed (e.g., the keyword count).
As the last step (line~14) Algorithm~\ref{algo:CC} executes the (partial) materialization procedure.

\textbf{Spatial Hierarchies Construction.}
In our proposed \emph{STTCube} the base level for the spatial hierarchies is the \textit{Location} present in the raw data, i.e., the longitude and latitude points.
Hence, we use Military Grid Reference System (MGRS) for grid-based hierarchy and  when building the semantic-based hierarchy, individual points are linked to the respective cities using the information in the available geographical taxonomy $\mathcal{G}$, or to a special member for points that link to unknown locations.
Therefore, this corresponds to the step function from \textit{Location} to \textit{City}.
The spatial taxonomy $\mathcal{G}$ is also used to generate the spatial hierarchy step functions for the higher levels.

\noindent\textbf{Textual Hierarchies Construction.}
The unstructured nature of the text makes it a challenging task to convert it into a dimension of a cube.
In Algorithm~\ref{algo:CC}, the \texttt{ProcessText} function (line~11) implements the following steps: (1) splits the text into individual words, (2) removes \textit{stop words}, and (3) converts the remaining words to their base form (e.g., ``works'' and  ``working'' have the same base form ``work'').
The final processed text is used to populate the \textit{Term} base-level in the textual dimension.
This implements the base step function in the textual hierarchies, and links every fact to one or more \textit{Terms}, hence it has an $n$$-$$n$ cardinality.
Moreover, while constructing the higher levels, using the semantic taxonomy $\mathcal{T}$ (e.g., WordNet), each \textit{STT object} is linked to one or more \textit{Themes}, and similarly for \textit{Topics}, and \textit{Concepts}.


\section{Experimental Evaluation}\label{sec:experiments}
\input{figures/QRTs_125M}

Now, we report on the performance of \emph{STTCube} analysis.
In particular, we compare the different materialization strategies for \emph{STTCube} and \emph{No STTCube (NC)} implementations, in terms of query response time (QRT) and storage cost.
\emph{NC} answers the queries by computing the query response from base data without constructing the \emph{STTCube}.
\emph{NC} stores \emph{pre-processed text}, lemmatized text after removal of stop and invalid words, along with the \emph{geographical location}, longitude and latitude point, and \emph{timestamp}.
Specifically, \emph{NC} uses user-defined functions for text (for retrieving individual terms) and location processing (e.g., identification of the city a particular longitude, latitude point belongs to) and built-in functions for timestamp. 
Further, \emph{NC} filters on location and timestamp for the queried \emph{area} and time and performs a series of joins, e.g., 4 joins for \emph{Concept} level, to retrieve information for the requested textual level.
Finally, it groups results on the textual and temporal columns, computes the \emph{STT measure values}, and performs the top-\emph{k} selection.
Also, we compare QRT and hierarchy construction time for different combinations of hierarchy schemes.
Moreover, we also report on the accuracy of \emph{PAM} and demonstrate the advantage in performance when compared to \emph{PEM}.
Lastly, we compare QRTs for different spatial and textual hierarchy schemes, showing that combinations of \emph{Grid-based spatial and Majority-based textual (GM)} hierarchy scheme achieves the fastest QRTs among all hierarchy combinations.

\para{Experimental Setup.}
We evaluate the \emph{STTCube} on a real-world Twitter dataset containing 125 million tweets collected over \emph{six} weeks.
Each tweet contains the tweet location, text, and time.
We implemented the \emph{STTCube} in a leading commercial RDBMS, called \textit{RDBMS-X} as we cannot disclose the name.
The proposed design is realized using a \textit{snowflake schema} to avoid redundancy in the dimension data.

We implemented the \textit{Pre-Processing (PP)} component, where the whole raw dataset is parsed and the relational tables are populated, in Java (v11).
All tests are run on a Windows Server machine with 2 Intel Xeon 2.50GHz CPUs and 16GB RAM.

We extracted the taxonomy for the spatial dimension from GeoNames~\cite{Geonames}.
For the \textit{City} level, we considered all the cities having $population > 1000$ and for the \textit{Region} level, we use administrative divisions information available in the GeoNames dataset.
We use the reverse geocoding process to find the city name for the \textit{Location} coordinates.

For the textual dimension, as a taxonomy for \textit{Terms}, \textit{Themes}, \textit{Topics}, and \textit{Concepts}, we use the widely used WordNet~\cite{wordnet}.
We use the direct \textit{HYPERNYM} link of WordNet to decide the parent member for a \textit{Term}, \textit{Theme}, and \textit{Topic}.
If a term is present in WordNet and has a super-class (\textit{HYPERNYM}) then the super-class becomes the parent of the term. Otherwise, it becomes its own parent (this avoid unbalanced hierarchies and UNKNOWN values in the hierarchy).
For text pre-processing --tokenization, lemmatization, and stop word removal-- we use the Stanford Core NLP library~\cite{manning2014stanford}.
We implemented the temporal dimension using the standard \emph{Date} and \emph{Time} functions supported in \emph{RDBMS-X}. 

\input{figures/QRTs_SR}
We implemented the \emph{semantic-based} and \emph{grid-based} hierarchy schemes for the spatial dimension, \emph{replication-based} and \emph{majority-based} hierarchy schemes for the textual dimension (Section~\ref{sec:SThierarchies}), and \emph{Date} hierarchy for the temporal dimension.

\textbf{Spatial, Textual, and Temporal Levels Members.} The base levels contain $40.1$ million unique \textit{Location Points} and $9.8$ million unique \textit{Terms}.
The GeoNames taxonomy contains $132K$ cities, divided into $4K$ administrative divisions (regions) for $247$ countries.
Among those, we have tweets for $104K$ cities, $3.8K$ regions, and $246$ distinct countries.
In the textual hierarchy, terms are grouped into $23.8K$ \textit{Themes}, $19.4K$ \textit{Topics}, and $17.6K$ \textit{Concepts}.
Furthermore, the temporal dimension spans over $37$ days.
Finally, for \emph{PAM} we materialize $K$$=$$31$ densest keywords.

\begin{table}[b!]
	\begin{center}
		\scriptsize
		\captionsetup{font=footnotesize, skip=0pt}
		\caption{Spatio-Textual-Temporal Queries}
		\label{tab:queries}
		\begin{tabular}
			{||c||l||}\hline	
			\textbf{Query} & \multicolumn{1}{c||}{\textbf{Description}} \\ \hline
			Q1 & Top-\textit{k} Dense/Volatile \textit{Terms} in a \textit{City} \emph{[time span]} \\ \hline
			Q2 & Top-\textit{k} Dense/Volatile \textit{Topics} in a \textit{City} [\emph{time span]}\\ \hline
			Q3 & Top-\textit{k} Dense/Volatile \textit{Concepts} in a \textit{Country} \emph{[time span]}\\ \hline
			Q4 & Top-\textit{k} Dense/Volatile \textit{Terms} in a \textit{Region} \emph{[time span]}\\ \hline
			Q5 & Top-\textit{k} Dense/Volatile \textit{Concepts} in a \textit{Region} \emph{[time span]}\\ \hline
			Q6 & Top-\textit{k} Dense/Volatile \textit{Themes} in a \textit{Region} \emph{[time span]}\\ \hline
			Q7 & Top-\textit{k} Dense/Volatile \textit{Terms} in a \textit{Country} \emph{[time span]}\\ \hline
			Q8 & Top-\textit{k} Dense/Volatile \textit{Terms} in a \textit{Country}\\
				 & Group by \textit{Region} \emph{[time span]}\\ \hline
			Q9 & Top-\textit{ALL} Dense/Volatile \textit{Topics} in a \textit{Country}\\
				 & Group by \textit{Region} \emph{[time span]}\\ \hline
		\end{tabular}
	\end{center}
\end{table}

We implemented \emph{Keyword Density}, \emph{Keyword Volatility}, \emph{Top-k Dense Keywords within an area}, and \textit{Top-k Volatile Keywords within an area} as prototypical \emph{STT measures}.
We compare (\textit{PEM}) and  (\textit{PAM}) strategies with the following three baselines.
\textbf{No STTCube (NC):} is the traditional RDBMS setup with all textual, spatial, and temporal functions implemented as built-in or user-defined functions.
\emph{NC} is the \emph{traditional} solution one would go for without the \emph{STTCube}.
\textbf{No Materialization (NM):} constructs the \emph{STTCube} and minimizes the storage cost by only materializing the base cuboid and computing all query responses from it.
\textbf{Full Materialization (FM):} minimizes the QRT by materializing every cuboid in the \emph{STTCube}. With this approach queries are answered through a lookup in the pre-computed cuboid.
\emph{These three baselines are at the extreme ends of the space-time trade-off and are usually infeasible for large datasets.}

\para{Queries.}
We perform experiments on five different sizes of datasets using nine different \emph{STT queries}.
Each \emph{STT query}, described in Table~\ref{tab:queries}, targets a different level of spatial, textual, and temporal granularity.
Each query requests either dense or volatile keywords with a range of time which is used for volatile but not used for dense keywords queries.
We execute each query ten times with randomly generated parameters for each method and report mean and standard deviation.

\para{Query Response Time.}
For \emph{Top-k Dense and Top-k Volatile Keywords within an area} measures, we compare the QRT of \textit{PEM} and \textit{PAM} methods with the \textit{NC}, \textit{NM}, and \textit{FM} baselines.
For \emph{Keyword Density} and \emph{Keyword Volatility}, no approximate solution is possible so we only compare \textit{PEM} with \textit{NC}, \textit{NM}, and \textit{FM}.
As the \emph{Majority-based} textual hierarchy scheme does not process \emph{Terms} (Section~\ref{sec:SThierarchies}), we only evaluate \emph{five} out of \emph{nine} queries requesting \emph{Theme}, \emph{Topic}, and \emph{Concept} for it (Figures~\ref{fig:QRT125M}e---\ref{fig:QRT125M}h).
Furthermore, we cannot evaluate \emph{PAM} for Q9 as no approximate solution is possible for it.
\input{figures/bar_charts}
\setlength{\textfloatsep}{1pt plus 0pt minus 0pt}
We plot results in Figures~\ref{fig:QRT125M}a---\ref{fig:QRT125M}h for 100\% (125M) of data, as the results are similar for smaller data sizes.
Each row in Figure~\ref{fig:QRT125M} shows the QRTs for one particular combination of spatio-textual hierarchy schemes.
Specifically, Figures~\ref{fig:QRT125M}a---\ref{fig:QRT125M}d show the QRTs for the Grid-based spatial and Replication-based textual (\textbf{GR}) hierarchy combination for all \emph{measures}. 
Similarly, Figures~\ref{fig:QRT125M}e---\ref{fig:QRT125M}h show QRTs for Grid-based spatial and Majority-based textual (\textbf{GM}) combinations\footnote[2]{Due to space constraint we have omitted the figures for Semantic-based spatial hierarchy (can be found in Appendices~\Cref{fig:QRT125M_App,fig:QRTsTrends_App}) as we observed similar results}. 
Figure~\ref{fig:QRT125M} has queries on the x-axis and QRTs in msec on the y-axis (note: log scale).
Figure~\ref{fig:QRT125M} confirms that \emph{NC} is $1$$-$$5$ orders of magnitude slower than \emph{NM}.
Specifically, regardless of the spatial hierarchy scheme, it is $1$$-$$2$ and $3$$-$$5$ orders of magnitude slower than \emph{NM} for \emph{Replication-based} and \emph{Majority-based} textual hierarchy, respectively.
The \emph{Majority-based} textual hierarchy scheme achieves faster QRTs because it does not process individual \emph{Terms} but directly links \emph{Theme} to the \emph{fact}, hence, drastically reducing the number of rows to process (from millions to thousands).
Furthermore, \textit{NM} is $1$$-$$4$ and $3$$-$$5$ orders of magnitude slower than \textit{PEM} and \emph{PAM}, respectively, for all measures and combinations of hierarchy schemes.
\emph{PEM} is on average six times slower than \textit{FM} which achieves its fast QRTs at the expense of a highly increased storage cost (Figure~\subref{fig:MemoryComparison}).
\textit{PAM} achieves \emph{near-optimal} QRTs because it materializes only \textit{the K densest keywords} in the cuboid, hence it has much fewer rows to process.
QRTs for \emph{Q9} for \emph{Top-\emph{k} Volatile Keyword within an area} and \emph{Top-\emph{k} Dense Keywords within an area} measures for all combinations of hierarchy schemes are the worst for \textit{PAM} (same as \textit{NM}) because it requests \emph{ALL} keywords\textquotesingle~densities instead of \emph{top-k} which cannot be computed from the approximate pre-aggregated information.
To generate a response for \emph{Q9}, we have to process all detail data directly from the base facts.
In comparison, \textit{PEM} and \textit{PAM} materialize a subset of views (also a subset of rows for \textit{PAM}) and use the pre-aggregated measure values in those views to efficiently generate a response for a query instead of processing base facts, thus improving the overall QRT.
\emph{NC} is the slowest of all ($1-5$ orders of magnitude slower than the slowest \emph{STTCube} \emph{NM}) because it has to process the complete dataset for computing each query response, and cannot take advantage of the \emph{STTCube} optimizations for \emph{STT measures}.
Among all the hierarchy scheme combinations, \emph{GM} has the fastest QRTs mainly because of \emph{Majority-based} which drastically reduces the row count by linking the \emph{Theme} directly to each \emph{Fact} instead of individual \emph{Terms}, whereas, \emph{GR} has the slowest QRTs due to \emph{Replication-based} having far more rows to process than \emph{Majority-based} textual hierarchy.
Furthermore, \emph{Grid} and \emph{Semantic-based} spatial hierarchies have similar QRTs. 


Figure~\ref{fig:QRTsTrends} shows the scalability of \textit{PEM} and \textit{PAM} over growing data sizes for different combinations of hierarchy schemes and confirms that the QRTs are almost constant as the data grows. 
This is because the sizes of materialized views do not increase a lot as the data grows.
Only new dimension members, e.g., new cities or topics, increase the size of materialized views, but only by a small fraction. 
Figures~\ref{fig:QRTsTrends}f---\ref{fig:QRTsTrends}j confirm that the \emph{GM} hierarchy combination results in the fastest QRTs, i.e., all QRTs $<$ 100 msec. 
On the contrary, Figures~\ref{fig:QRTsTrends}a---\ref{fig:QRTsTrends}e show that \emph{GR} yields the slowest QRTs, with QRT as high as 400 msec. 
Figure~\ref{fig:QRTsTrends} confirms that \emph{PAM} consistently achieves the fastest QRTs (mostly $<$ 100 msec with few a bit over 100 msec) regardless of hierarchy schemes. 
Figure~\ref{fig:QRTsTrends} shows that \textit{PEM} and \textit{PAM} scale linearly w.r.t. data size.

\para{Storage Cost.}
\label{sec:storage_cost}
We now compare the storage cost for \textit{FM}, \textit{PEM}, \textit{PAM}, and \textit{NM}.
We do not compare \emph{NC}\textquotesingle s storage cost because it does not construct STTCube, and hence does not materialize anything.
We only show the storage cost for up to 20 million because \emph{FM} takes an unfeasible amount of time (shown in Figure~\subref{fig:PPCC}) while for the other methods and over the larger datasets we observe the same trend.
We use the number of rows in a view as its storage cost.
The base cube\textquotesingle s storage cost is always needed.
Besides that, every additional materialized view adds to the storage cost, as displayed in Figure~\subref{fig:MemoryComparison}, that shows the storage cost of \textit{NM}, \textit{PAM}, \textit{PEM}, and \textit{FM} over growing data sizes.
The materialization of the STTCube using \textit{PEM} and \textit{PAM} only adds 13\%  and 0.1\% to the storage cost of the base cube, respectively.
Whereas, using \textit{FM} increases the storage cost by more than an order of magnitude. 
\textit{PEM} reduces the storage cost by only materializing a subset of views (four views) and still achieves 2-5 orders of magnitude improvement in QRT (Figures~\ref{fig:QRT125M}).
\textit{PAM} further reduces the storage cost by only materializing a subset of rows in a view (top-{\textit{k}}) and gains an additional order of magnitude improvement in QRT.
On the other hand, \textit{FM} materializes all views in a cube, i.e., 500 ($5\times5\times5\times4$) views in our case, which makes the view materialization storage cost much higher (one order of magnitude) than the base cube itself, as shown in Figure~\subref{fig:MemoryComparison}.
Figure~\subref{fig:MemoryComparison} confirms that our proposed methods \textit{PEM} and \textit{PAM} reduce the storage cost between 97\% and 99.9\% compared to \textit{FM}.

\begin{figure*}[t!]
	\tiny
	\centering
	\captionsetup{font=footnotesize, skip=0pt}

\pgfplotsset{
	box plot width/.initial=1em,
	box plot/.style={
		/pgfplots/.cd,
		black,
		only marks,
		mark=-,
		mark size=\pgfkeysvalueof{/pgfplots/box plot width},
		/pgfplots/error bars/.cd,
		y dir=plus,
		y explicit,
	},
	box plot box/.style={
		/pgfplots/error bars/draw error bar/.code 2 args={%
			\draw  ##1 -- ++(\pgfkeysvalueof{/pgfplots/box plot width},0pt) |- ##2 -- ++(-\pgfkeysvalueof{/pgfplots/box plot width},0pt) |- ##1 -- cycle;
		},
		/pgfplots/table/.cd,
		y index=2,
		y error expr={\thisrowno{3}-\thisrowno{2}},
		/pgfplots/box plot
	},
	box plot top whisker/.style={
		/pgfplots/error bars/draw error bar/.code 2 args={%
			\pgfkeysgetvalue{/pgfplots/error bars/error mark}%
			{\pgfplotserrorbarsmark}%
			\pgfkeysgetvalue{/pgfplots/error bars/error mark options}%
			{\pgfplotserrorbarsmarkopts}%
			\path ##1 -- ##2;
		},
		/pgfplots/table/.cd,
		y index=4,
		y error expr={\thisrowno{2}-\thisrowno{4}},
		/pgfplots/box plot
	},
	box plot bottom whisker/.style={
		/pgfplots/error bars/draw error bar/.code 2 args={%
			\pgfkeysgetvalue{/pgfplots/error bars/error mark}%
			{\pgfplotserrorbarsmark}%
			\pgfkeysgetvalue{/pgfplots/error bars/error mark options}%
			{\pgfplotserrorbarsmarkopts}%
			\path ##1 -- ##2;
		},
		/pgfplots/table/.cd,
		y index=5,
		y error expr={\thisrowno{3}-\thisrowno{5}},
		/pgfplots/box plot
	},
	box plot median/.style={
		/pgfplots/box plot
	}
}
	\subfloat[PAM\textquotesingle s Accuracy]{
				\begin{tabular}
					{||c||c|c|c|c|c||}
					\hline
					\multirow{2}{*}{\textbf{Query}} & \multicolumn{5}{c||}{Data Size in Millions} \\ \cline{2-6}
					
					& 25 & 50 & 75 & 100 & 125 \\ \hline
					Q1 & 100.0 & 100.0 & 100.0 & 100.0 & 100.0 \\ \hline
					Q2 & 100.0 & 100.0 & 100.0 & 100.0 & 100.0 \\ \hline
					Q3 & 100.0 & 100.0 & 90.0 & 95.0 & 90.0\\ \hline
					Q4 & 92.3 & 100.0 & 100.0 & 100.0 & 100.0\\ \hline
					Q5 & 100.0 & 100.0 & 100.0 & 100.0 & 100.0 \\ \hline
					Q6 & 100.0 & 100.0 & 100.0 & 100.0 & 100.0\\ \hline
					Q7 & 93.3 & 96.7 & 90.0 & 93.3 & 93.3\\ \hline
					Q8 & 100.0 & 100.0 & 100.0 & 100.0 & 100.0\\ \hline
				\end{tabular} \label{tab:accuracy}}\qquad
%
%
%
%
	\subfloat[STTOLAP Operations\textquotesingle \,QRTs]{
	\begin{tikzpicture}
		\begin{axis}[
				height=3.5cm,
				ymode = log,
				xtick pos=left,
				ytick pos=left,
				ylabel={QRT (ms)},
				xlabel={STTOLAP Operations},
				ylabel style={yshift = {-5}},
				legend entries={NM, PEM, PAM, FM},
				ytick={10,100,1000, 10000, 100000, 1000000},
				xtick={1,...,7},
				xticklabels={Start, RU, RU, D, S, DD, DD},
				legend pos=outer north east,
				legend columns = 1,yminorticks=false
				]
				
				\addplot coordinates {(1,880654.0) (2,858564.0) (3,873685.0) (4,220015.0) (5,219898.0) (6,965695.0) (7,849125.0)};
				
				\addplot coordinates {(1,1262.0)(2,4824.0)(3,6248.0)(4,58.0)(5,32.0)(6,265.0)(7,1204.0)};
				
				\addplot coordinates {(1,127.0)(2,766.0)(3,2219.0)(4,41.0)(5,47.0)(6,15.0)(7,141.0)};
				
				\addplot coordinates {(1,123.0) (2,516.0) (3,915.0) (4,32.0) (5,32.0) (6,16.0) (7,64.0)};
			\end{axis}
	\end{tikzpicture}
\label{fig:STTOLAP}
}\qquad
	\subfloat[Materialized-K Vs QRT]{
	\begin{tikzpicture}
		\begin{axis} 
			[enlarge x limits=0.1,box plot width=0.5em, ymode=log, xtick=data, xticklabels = {10,20,50,100,200,500,1000}, xlabel={Materialized K}, ylabel={QRTs (msec)}, ymin=1, ytick={10,100,1000, 10000, 100000, 1000000},xtick pos=left,
			ytick pos=left,
			height=3.5cm,yminorticks=false
			]
			
			\addplot [box plot median] table {testdata.dat};
			
			\addplot [box plot box] table {testdata.dat};
			
			\addplot [box plot top whisker] table {testdata.dat};
			
			\addplot [box plot bottom whisker] table {testdata.dat};
		\end{axis}
\end{tikzpicture}
\label{fig:TopKAnalysis}}
\end{figure*}
\para{Views Selection for Materialization.}
Our proposed methods \textit{PEM} and \textit{PAM} are partial materialization methods that materialize only a subset of the cuboids.
Hence, an important trade-off to be understood is between the number of cuboids to materialize, the corresponding storage cost, and the gain in query response time achieved.
We empirically evaluate the benefit gained (improvements in QRT for all dependent cells which can be answered using this view) against the cost of materializing the view (Algorithm~\ref{algo:GHM}).
We consider the base cube as a necessary view to be materialized and consider its benefit as zero.
Figure~\subref{fig:ViewsVsBenefit} shows that materializing three cuboids (\textit{(Day, City, Term)}, \textit{(Day, Location, Theme)}, and \textit{(Day, Region, Term)}) on top of the base cube gain the most benefit after which we do not get a significant advantage of materializing further cuboids.
The reason is that the materialized cuboids are already small enough, so the benefit of materializing any descendant cuboid is small.
Hence, materializing 4 cuboids represents the best trade-off between QRT and storage cost.

\para{Pre-Processing and Cube Construction.}
Here, we report the time for the construction of  \emph{STTCube}.
Construction of an STTCube is divided into two steps: 1) \textit{Pre-Processing (PP)} of base facts (\emph{spatio-textual-temporal objects}) and population of the relational tables and 2) materialization of views.
Further, the materialization of views can be done either using \textit{FM}, \textit{PEM}, or \textit{PAM}.
In Figure~\subref{fig:PPCC}, we have data sizes on the x-axis and time in minutes on the y-axis (note: log scale).
\textit{FM} is the most time consuming among all and adds significant overhead on top of \emph{PP} time and does not scale. 
On the contrary, \emph{PEM} and \emph{PAM} time is negligible compared to the \emph{FM} time.
Hence, with \emph{PEM} and \emph{PAM} \emph{STTCube} construction time scales linearly.
To evaluate \emph{STTCube's} ability to handle updates, we performed several updates of 4M tweets each (\emph{PP\_INC} line in Figure~\subref{fig:PPCC}).
Experiments confirm that STTCube handles updates efficiently by only processing the new \emph{STT objects} and updating the respective dimensions and measures.

Furthermore, we compare the different hierarchy schemes w.r.t. their construction time.
Figure~\subref{fig:PPCC_ALL} shows the hierarchies' construction time for different hierarchy schemes.
It is evident from Figure~\subref{fig:PPCC_ALL} that, among all, the \emph{Replication-based} textual hierarchy scheme takes the longest to construct because for each single \emph{spatio-textual-temporal} object it has to process each individual \emph{Term} and construct hierarchy for it.
Whereas, for all other schemes, for each spatio-textual-temporal object only one hierarchy instance is processed.
Figure~\subref{fig:PPCC_ALL} confirms that all of the hierarchy schemes are constructed in linear time w.r.t. data size, allowing STTCube to support multiple hierarchy schemes.

\para{Accuracy.}
Given that \textit{PAM} efficiently computes the approximate measure values, it becomes necessary to evaluate its accuracy.
To evaluate the accuracy of \textit{PAM}, we use \textit{NM}\textquotesingle s results as ground truth.
Our evaluation result in Table~\subref{tab:accuracy} confirms that it achieves high accuracy.
Specifically, it is 100\% for 6 out of 8 queries, and 90-97\% for 2.
Queries with 90-97\% accuracy request as many keywords as are materialized and the risk of having wrong results near the border (bottom of the top-\textit{k} list) is higher.

\para{QRT of STTOLAP Operators.}
Our proposed materialization strategies (\emph{PEM and PAM}) improves the QRTs for \emph{STTOLAP operators}.
To demonstrate this, we perform a series of STTOLAP operations and measure their QRT for different materialization strategies.
Figure~\subref{fig:STTOLAP} shows the QRTs for multiple \emph{STTOLAP operations} for different materialization strategies.
We have STTOLAP operators on the x-axis (RU, D, S, and DD represents STT Roll Up, Dice, Slice, and Drill Down operators, respectively) on QRT in msec on the y-axis.
It is evident that \emph{NM} is on average 3-5 orders of magnitude slower than \emph{PEM} which is one order of magnitude slower than \emph{PAM}.
Furthermore, \emph{PAM} achieves near-optimal QRTs, just a fraction higher than \emph{FM}.
These experiments confirm that \emph{STTCube\textquotesingle s} materialization methods (\emph{PEM and PAM}) improves \emph{STTOLAP operators\textquotesingle} QRTs by materializing only a subset of cuboids.

\para{Top-K Value Estimation.}
Here, we study the relationship between QRT and the value of materialized $K$.
We create \emph{seven} different \emph{STTCube} materialization versions using 10, 20, 50, 100, 200, 500, and 1000 as the value of $K$.
Next, we use the Gamma distribution to generate 100 random numbers, to be used as top-\emph{k} values, in the range of 1 and 1000.
We chose the Gamma distribution because it resembles a common long-tail distribution for top-\emph{k} values.
We execute each query for all the 100 generated top-\emph{k} values over all \emph{seven} materialization versions.
Figure~\subref{fig:TopKAnalysis} shows the QRT for all queries over different materialization versions.
For $K$$=$$10$ and $20$ the median value is the same as the box top, hence not visible in the plot.
It is evident from Figure~\subref{fig:TopKAnalysis} that a larger value of materialized $K$ achieves faster QRTs (lower median value) because almost all the queries are answered using the pre-computed measure values. 
But, in the case of smaller $K$, all the queries requesting $k$$>$$K$ need to be answered using the non-pre-computed measure values from the base cuboid.
Hence, resulting in slower QRTs (higher median value).
A larger value of $K$ such as $1000$ is not recommended because 1) there will be very few queries requesting a larger top-\emph{k} and 2) it will require more storage cost (Figure~\subref{fig:MemoryComparison}).
Specifically, between $K$$=$$50$ and $100$ and $K$$=$$100$ and $200$ QRT decrease by 35\% and 0\% but storage increase 250\% and 200\%, respectively. 
Hence, these experiments confirm that choosing a value between 20---50 for $K$ in our current experiments settings is a near-optimal choice.
\section{Conclusion and Future Work}\label{sec:conclusion}
In this paper, we defined and formalized the \emph{Spatio-Textual-Temp-oral Cube (STTCube)} structure to effectively perform \emph{STTCube analytics}.
\emph{We introduced STT hierarchies, spatio-textual and spatio-textual-temporal measures, and STTOLAP operators to analyze STT data together.} 
For efficient, exact and approximate, computation of \emph{STT measures}, we proposed a pre-aggregation framework able to provide faster response times by requiring a controlled amount of extra storage to store pre-computed measure values.
We observed how the partial materialization provides 1 to 5 orders of magnitude reduction in query response time, with between 97\% and 99.9\% reduced storage cost compared to full materialization techniques.
Moreover, the approximate materialization provides accuracy between 90\% and 100\%, while requiring considerably less space compared to no materialization techniques.
In future work, we plan to enhance STTCube with additional \emph{STT measures} and distributed implementation. 

\bibliographystyle{ACM-Reference-Format}
\bibliography{bibliography}


\begin{thebibliography}{38}


\ifx \showCODEN    \undefined \def \showCODEN     #1{\unskip}     \fi
\ifx \showDOI      \undefined \def \showDOI       #1{#1}\fi
\ifx \showISBNx    \undefined \def \showISBNx     #1{\unskip}     \fi
\ifx \showISBNxiii \undefined \def \showISBNxiii  #1{\unskip}     \fi
\ifx \showISSN     \undefined \def \showISSN      #1{\unskip}     \fi
\ifx \showLCCN     \undefined \def \showLCCN      #1{\unskip}     \fi
\ifx \shownote     \undefined \def \shownote      #1{#1}          \fi
\ifx \showarticletitle \undefined \def \showarticletitle #1{#1}   \fi
\ifx \showURL      \undefined \def \showURL       {\relax}        \fi
\providecommand\bibfield[2]{#2}
\providecommand\bibinfo[2]{#2}
\providecommand\natexlab[1]{#1}
\providecommand\showeprint[2][]{arXiv:#2}

\bibitem[\protect\citeauthoryear{??}{Geo}{2020}]%
        {Geonames}
 \bibinfo{year}{2020}\natexlab{}.
\newblock \bibinfo{title}{{GeoNames}}.
\newblock \bibinfo{howpublished}{\url{http://download.geonames.org/}}.
\newblock
\newblock
\shownote{Accessed: 2020-09-09.}


\bibitem[\protect\citeauthoryear{Almaslukh, Magdy, Aly, Mokbel, Elnikety, He,
  Nath, and Aref}{Almaslukh et~al\mbox{.}}{2019}]%
        {GeoTrend+}
\bibfield{author}{\bibinfo{person}{A. Almaslukh}, \bibinfo{person}{A. Magdy},
  \bibinfo{person}{A.~M. Aly}, \bibinfo{person}{M.~F. Mokbel},
  \bibinfo{person}{S. Elnikety}, \bibinfo{person}{Y. He}, \bibinfo{person}{S.
  Nath}, {and} \bibinfo{person}{W.~G. Aref}.} \bibinfo{year}{2019}\natexlab{}.
\newblock \showarticletitle{Local trend discovery on real-time microblogs with
  uncertain locations in tight memory environments}.
\newblock \bibinfo{journal}{\emph{GeoInformatica}} (\bibinfo{year}{2019}).
\newblock


\bibitem[\protect\citeauthoryear{{Azabou}, {Khrouf}, {Feki}, {Soulé-Dupuy},
  and {Vallès}}{{Azabou} et~al\mbox{.}}{2016}]%
        {newTextOLAP}
\bibfield{author}{\bibinfo{person}{M. {Azabou}}, \bibinfo{person}{K. {Khrouf}},
  \bibinfo{person}{J. {Feki}}, \bibinfo{person}{C. {Soulé-Dupuy}}, {and}
  \bibinfo{person}{N. {Vallès}}.} \bibinfo{year}{2016}\natexlab{}.
\newblock \showarticletitle{Analyzing textual documents with new OLAP
  operators}.
\newblock \bibinfo{journal}{\emph{AICCSA}} (\bibinfo{year}{2016}).
\newblock


\bibitem[\protect\citeauthoryear{Cao, Chen, Cong, Jensen, Qu, Skovsgaard, Wu,
  and Yiu}{Cao et~al\mbox{.}}{2012}]%
        {SpatialQuery}
\bibfield{author}{\bibinfo{person}{X. Cao}, \bibinfo{person}{L. Chen},
  \bibinfo{person}{G. Cong}, \bibinfo{person}{C.~S. Jensen},
  \bibinfo{person}{Q. Qu}, \bibinfo{person}{A. Skovsgaard}, \bibinfo{person}{D.
  Wu}, {and} \bibinfo{person}{M.~L. Yiu}.} \bibinfo{year}{2012}\natexlab{}.
\newblock \showarticletitle{Spatial Keyword Querying}.
\newblock \bibinfo{journal}{\emph{ER}} (\bibinfo{year}{2012}).
\newblock


\bibitem[\protect\citeauthoryear{Chen, Cong, Jensen, and Wu}{Chen
  et~al\mbox{.}}{2013}]%
        {ExperimentalEvaluation}
\bibfield{author}{\bibinfo{person}{L. Chen}, \bibinfo{person}{G. Cong},
  \bibinfo{person}{C.~S. Jensen}, {and} \bibinfo{person}{D. Wu}.}
  \bibinfo{year}{2013}\natexlab{}.
\newblock \showarticletitle{Spatial Keyword Query Processing: An Experimental
  Evaluation}.
\newblock \bibinfo{journal}{\emph{PVLDB}} (\bibinfo{year}{2013}).
\newblock


\bibitem[\protect\citeauthoryear{Chen, Dong, Han, Wah, and Wang}{Chen
  et~al\mbox{.}}{2002}]%
        {TimeSeries}
\bibfield{author}{\bibinfo{person}{Y. Chen}, \bibinfo{person}{G. Dong},
  \bibinfo{person}{J. Han}, \bibinfo{person}{B.~W. Wah}, {and}
  \bibinfo{person}{J. Wang}.} \bibinfo{year}{2002}\natexlab{}.
\newblock \showarticletitle{Multi-dimensional Regression Analysis of
  Time-series Data Streams}.
\newblock \bibinfo{journal}{\emph{VLDB}} (\bibinfo{year}{2002}).
\newblock


\bibitem[\protect\citeauthoryear{Chouder, Rizzi, and Chalal}{Chouder
  et~al\mbox{.}}{2019}]%
        {EXODuS}
\bibfield{author}{\bibinfo{person}{M.~L. Chouder}, \bibinfo{person}{S. Rizzi},
  {and} \bibinfo{person}{R. Chalal}.} \bibinfo{year}{2019}\natexlab{}.
\newblock \showarticletitle{Exploratory OLAP over Doc Stores}.
\newblock \bibinfo{journal}{\emph{IS}} (\bibinfo{year}{2019}).
\newblock


\bibitem[\protect\citeauthoryear{{Cong}, {Feng}, and {Zhao}}{{Cong}
  et~al\mbox{.}}{2016}]%
        {GeoTextualChallenges}
\bibfield{author}{\bibinfo{person}{G. {Cong}}, \bibinfo{person}{K. {Feng}},
  {and} \bibinfo{person}{K. {Zhao}}.} \bibinfo{year}{2016}\natexlab{}.
\newblock \showarticletitle{Querying and mining geo-textual data for
  exploration: Challenges and opportunities}.
\newblock \bibinfo{journal}{\emph{ICDEW}} (\bibinfo{year}{2016}).
\newblock


\bibitem[\protect\citeauthoryear{Cong and Jensen}{Cong and Jensen}{2016}]%
        {QueryingGeoTextual}
\bibfield{author}{\bibinfo{person}{G. Cong} {and} \bibinfo{person}{C.~S.
  Jensen}.} \bibinfo{year}{2016}\natexlab{}.
\newblock \showarticletitle{Spatial Keyword Queries and Beyond}.
\newblock \bibinfo{journal}{\emph{SIGMOD}} (\bibinfo{year}{2016}).
\newblock


\bibitem[\protect\citeauthoryear{{Ding}, {Zhao}, {Lin}, {Han}, {Zhai},
  {Srivastava}, and {Oza}}{{Ding} et~al\mbox{.}}{2011}]%
        {TextCubeTopKCells}
\bibfield{author}{\bibinfo{person}{B. {Ding}}, \bibinfo{person}{B. {Zhao}},
  \bibinfo{person}{C.~X. {Lin}}, \bibinfo{person}{J. {Han}},
  \bibinfo{person}{C. {Zhai}}, \bibinfo{person}{A. {Srivastava}}, {and}
  \bibinfo{person}{N.~C. {Oza}}.} \bibinfo{year}{2011}\natexlab{}.
\newblock \showarticletitle{Efficient Keyword-Based Search for Top-K Cells in
  Text Cube}.
\newblock \bibinfo{journal}{\emph{TKDE}} (\bibinfo{year}{2011}).
\newblock


\bibitem[\protect\citeauthoryear{Fellbaum}{Fellbaum}{1998}]%
        {wordnet}
\bibfield{author}{\bibinfo{person}{C. Fellbaum}.}
  \bibinfo{year}{1998}\natexlab{}.
\newblock \showarticletitle{WordNet: An Electronic Lexical Database}.
\newblock \bibinfo{journal}{\emph{MIT Press}} (\bibinfo{year}{1998}).
\newblock


\bibitem[\protect\citeauthoryear{Feng, Zhang, Zhang, Han, Wang, Aggarwal, and
  Huang}{Feng et~al\mbox{.}}{2015}]%
        {StreamCube}
\bibfield{author}{\bibinfo{person}{W. Feng}, \bibinfo{person}{C. Zhang},
  \bibinfo{person}{W. Zhang}, \bibinfo{person}{J. Han}, \bibinfo{person}{J.
  Wang}, \bibinfo{person}{C. Aggarwal}, {and} \bibinfo{person}{J. Huang}.}
  \bibinfo{year}{2015}\natexlab{}.
\newblock \showarticletitle{STREAMCUBE: Hierarchical spatio-temporal hashtag
  clustering for event exploration over the Twitter stream}.
\newblock \bibinfo{journal}{\emph{ICDE}} (\bibinfo{year}{2015}).
\newblock


\bibitem[\protect\citeauthoryear{{Gray}, {Bosworth}, {Lyaman}, and
  {Pirahesh}}{{Gray} et~al\mbox{.}}{1996}]%
        {JimGray}
\bibfield{author}{\bibinfo{person}{J. {Gray}}, \bibinfo{person}{A. {Bosworth}},
  \bibinfo{person}{A. {Lyaman}}, {and} \bibinfo{person}{H. {Pirahesh}}.}
  \bibinfo{year}{1996}\natexlab{}.
\newblock \showarticletitle{Data cube: a relational aggregation operator
  generalizing GROUP-BY, CROSS-TAB, and SUB-TOTALS}.
\newblock \bibinfo{journal}{\emph{ICDE}} (\bibinfo{year}{1996}).
\newblock


\bibitem[\protect\citeauthoryear{Gür, Pedersen, Zimanyi, and Hose}{Gür
  et~al\mbox{.}}{2017}]%
        {sweb}
\bibfield{author}{\bibinfo{person}{N. Gür}, \bibinfo{person}{T.~B. Pedersen},
  \bibinfo{person}{E. Zimanyi}, {and} \bibinfo{person}{K. Hose}.}
  \bibinfo{year}{2017}\natexlab{}.
\newblock \showarticletitle{A foundation for spatial data warehouses on the
  Semantic Web}.
\newblock \bibinfo{journal}{\emph{Semantic Web}} (\bibinfo{year}{2017}).
\newblock


\bibitem[\protect\citeauthoryear{Han, Koperski, and Stefanovic}{Han
  et~al\mbox{.}}{1997}]%
        {GeoMiner}
\bibfield{author}{\bibinfo{person}{J. Han}, \bibinfo{person}{K. Koperski},
  {and} \bibinfo{person}{N. Stefanovic}.} \bibinfo{year}{1997}\natexlab{}.
\newblock \showarticletitle{GeoMiner: A System Prototype for Spatial Data
  Mining}.
\newblock \bibinfo{journal}{\emph{SIGMOD}} (\bibinfo{year}{1997}).
\newblock


\bibitem[\protect\citeauthoryear{Harinarayan, Rajaraman, and
  Ullman}{Harinarayan et~al\mbox{.}}{1996}]%
        {ImpCubeEff}
\bibfield{author}{\bibinfo{person}{V. Harinarayan}, \bibinfo{person}{A.
  Rajaraman}, {and} \bibinfo{person}{J.~D. Ullman}.}
  \bibinfo{year}{1996}\natexlab{}.
\newblock \showarticletitle{Implementing data cubes efficiently}.
\newblock \bibinfo{journal}{\emph{SIGMOD}} (\bibinfo{year}{1996}).
\newblock


\bibitem[\protect\citeauthoryear{Jayachandran, Tunga, Kamat, and
  Nandi}{Jayachandran et~al\mbox{.}}{2014}]%
        {DICE}
\bibfield{author}{\bibinfo{person}{P. Jayachandran}, \bibinfo{person}{K.
  Tunga}, \bibinfo{person}{N. Kamat}, {and} \bibinfo{person}{A. Nandi}.}
  \bibinfo{year}{2014}\natexlab{}.
\newblock \showarticletitle{Combining User Interaction, Speculative Query
  Execution and Sampling in the DICE System}.
\newblock \bibinfo{journal}{\emph{ICDE}} (\bibinfo{year}{2014}).
\newblock


\bibitem[\protect\citeauthoryear{Jensen, Pedersen, and Thomsen}{Jensen
  et~al\mbox{.}}{2010}]%
        {dwbook}
\bibfield{author}{\bibinfo{person}{C.~S. Jensen}, \bibinfo{person}{T.~B.
  Pedersen}, {and} \bibinfo{person}{C. Thomsen}.}
  \bibinfo{year}{2010}\natexlab{}.
\newblock \bibinfo{booktitle}{\emph{Multidimensional Databases and Data
  Warehousing}}.
\newblock \bibinfo{publisher}{Morgan \& Claypool Publishers}.
\newblock


\bibitem[\protect\citeauthoryear{Kenney and Keeping}{Kenney and
  Keeping}{1962}]%
        {RateOfChange}
\bibfield{author}{\bibinfo{person}{J.~F. Kenney} {and} \bibinfo{person}{E.~S.
  Keeping}.} \bibinfo{year}{1962}\natexlab{}.
\newblock \showarticletitle{Mathematics of Statistics, Part 1, chapter 15}.
\newblock \bibinfo{journal}{\emph{van Nostrand}} (\bibinfo{year}{1962}).
\newblock


\bibitem[\protect\citeauthoryear{Knijff, Frasincar, and Hogenboom}{Knijff
  et~al\mbox{.}}{2013}]%
        {taxonomy}
\bibfield{author}{\bibinfo{person}{J.~D. Knijff}, \bibinfo{person}{F.
  Frasincar}, {and} \bibinfo{person}{F. Hogenboom}.}
  \bibinfo{year}{2013}\natexlab{}.
\newblock \showarticletitle{Domain taxonomy learning from text: The subsumption
  method versus hierarchical clustering}.
\newblock \bibinfo{journal}{\emph{DKE}} (\bibinfo{year}{2013}).
\newblock


\bibitem[\protect\citeauthoryear{Lieberman, Samet, Sankaranarayanan, and
  Sperling}{Lieberman et~al\mbox{.}}{2007}]%
        {STEWARD}
\bibfield{author}{\bibinfo{person}{M.~D. Lieberman}, \bibinfo{person}{H.
  Samet}, \bibinfo{person}{J. Sankaranarayanan}, {and} \bibinfo{person}{J.
  Sperling}.} \bibinfo{year}{2007}\natexlab{}.
\newblock \showarticletitle{STEWARD: Architecture of a Spatio-textual Search
  Engine}.
\newblock \bibinfo{journal}{\emph{GIS}} (\bibinfo{year}{2007}).
\newblock


\bibitem[\protect\citeauthoryear{Lin, Ding, Han, Zhu, and Zhao}{Lin
  et~al\mbox{.}}{2008}]%
        {TextCube}
\bibfield{author}{\bibinfo{person}{C.~X. Lin}, \bibinfo{person}{B. Ding},
  \bibinfo{person}{J. Han}, \bibinfo{person}{F. Zhu}, {and} \bibinfo{person}{B.
  Zhao}.} \bibinfo{year}{2008}\natexlab{}.
\newblock \showarticletitle{Text Cube: Computing IR Measures for
  Multidimensional Text Database Analysis}.
\newblock \bibinfo{journal}{\emph{ICDM}} (\bibinfo{year}{2008}).
\newblock


\bibitem[\protect\citeauthoryear{Lins, Klosowski, and Scheidegger}{Lins
  et~al\mbox{.}}{2013}]%
        {Nanocubes}
\bibfield{author}{\bibinfo{person}{L. Lins}, \bibinfo{person}{J.~T. Klosowski},
  {and} \bibinfo{person}{C.~E. Scheidegger}.} \bibinfo{year}{2013}\natexlab{}.
\newblock \showarticletitle{Nanocubes for real-time exploration of
  spatiotemporal datasets}.
\newblock \bibinfo{journal}{\emph{TVCG}} (\bibinfo{year}{2013}).
\newblock


\bibitem[\protect\citeauthoryear{Liu, Tang, Hancock, Han, Song, Xu, and
  Pokorny}{Liu et~al\mbox{.}}{2013}]%
        {TextCubeApproach}
\bibfield{author}{\bibinfo{person}{X. Liu}, \bibinfo{person}{K. Tang},
  \bibinfo{person}{J. Hancock}, \bibinfo{person}{J. Han}, \bibinfo{person}{M.
  Song}, \bibinfo{person}{R. Xu}, {and} \bibinfo{person}{B. Pokorny}.}
  \bibinfo{year}{2013}\natexlab{}.
\newblock \showarticletitle{A Text Cube Approach to Human Social and Cultural
  Behavior in the Twitter Stream}.
\newblock \bibinfo{journal}{\emph{SBP}} (\bibinfo{year}{2013}).
\newblock


\bibitem[\protect\citeauthoryear{Magdy, Abdelhafeez, Kang, Ong, and
  Mokbel}{Magdy et~al\mbox{.}}{2020}]%
        {microblogsSurvey}
\bibfield{author}{\bibinfo{person}{A. Magdy}, \bibinfo{person}{L. Abdelhafeez},
  \bibinfo{person}{Y. Kang}, \bibinfo{person}{E. Ong}, {and}
  \bibinfo{person}{M.F. Mokbel}.} \bibinfo{year}{2020}\natexlab{}.
\newblock \showarticletitle{Microblogs data management: a survey}.
\newblock \bibinfo{journal}{\emph{The VLDB Journal}} (\bibinfo{year}{2020}).
\newblock


\bibitem[\protect\citeauthoryear{Manning, Surdeanu, Bauer, Finkel, Bethard, and
  McClosky}{Manning et~al\mbox{.}}{2014}]%
        {manning2014stanford}
\bibfield{author}{\bibinfo{person}{C. Manning}, \bibinfo{person}{M. Surdeanu},
  \bibinfo{person}{J. Bauer}, \bibinfo{person}{J. Finkel}, \bibinfo{person}{S.
  Bethard}, {and} \bibinfo{person}{D. McClosky}.}
  \bibinfo{year}{2014}\natexlab{}.
\newblock \showarticletitle{The Stanford CoreNLP natural language processing
  toolkit}.
\newblock \bibinfo{journal}{\emph{ACL}} (\bibinfo{year}{2014}).
\newblock


\bibitem[\protect\citeauthoryear{Othman, Belkaroui, and Faiz}{Othman
  et~al\mbox{.}}{2016}]%
        {userOpinions}
\bibfield{author}{\bibinfo{person}{R. Othman}, \bibinfo{person}{R. Belkaroui},
  {and} \bibinfo{person}{R. Faiz}.} \bibinfo{year}{2016}\natexlab{}.
\newblock \showarticletitle{Customer Opinion Summarization Based on Twitter
  Conversations}.
\newblock \bibinfo{journal}{\emph{WIMS}} (\bibinfo{year}{2016}).
\newblock


\bibitem[\protect\citeauthoryear{Pat and Kanza}{Pat and Kanza}{2017}]%
        {Waldo}
\bibfield{author}{\bibinfo{person}{B. Pat} {and} \bibinfo{person}{Y. Kanza}.}
  \bibinfo{year}{2017}\natexlab{}.
\newblock \showarticletitle{Where's Waldo?: Geosocial Search over Myriad
  Geotagged Posts}.
\newblock \bibinfo{journal}{\emph{SIGSPATIAL}} (\bibinfo{year}{2017}).
\newblock


\bibitem[\protect\citeauthoryear{P{\'e}rez-Mart\'{\i}nez, Berlanga-Llavori,
  Aramburu-Cabo, and Pedersen}{P{\'e}rez-Mart\'{\i}nez et~al\mbox{.}}{2008}]%
        {ContextualizedWarehouse}
\bibfield{author}{\bibinfo{person}{J.~M. P{\'e}rez-Mart\'{\i}nez},
  \bibinfo{person}{R. Berlanga-Llavori}, \bibinfo{person}{M.~J. Aramburu-Cabo},
  {and} \bibinfo{person}{T.~B. Pedersen}.} \bibinfo{year}{2008}\natexlab{}.
\newblock \showarticletitle{Contextualizing Data Warehouses with Documents}.
\newblock \bibinfo{journal}{\emph{DSS}} (\bibinfo{year}{2008}).
\newblock


\bibitem[\protect\citeauthoryear{Sankaranarayanan, Samet, Teitler, Lieberman,
  and Sperling}{Sankaranarayanan et~al\mbox{.}}{2009}]%
        {TwitterSand}
\bibfield{author}{\bibinfo{person}{J. Sankaranarayanan}, \bibinfo{person}{H.
  Samet}, \bibinfo{person}{B.~E. Teitler}, \bibinfo{person}{M.~D. Lieberman},
  {and} \bibinfo{person}{J. Sperling}.} \bibinfo{year}{2009}\natexlab{}.
\newblock \showarticletitle{TwitterStand: News in Tweets}.
\newblock \bibinfo{journal}{\emph{SIGSPATIAL}} (\bibinfo{year}{2009}).
\newblock


\bibitem[\protect\citeauthoryear{Simitsis, Baid, Sismanis, and
  Reinwald}{Simitsis et~al\mbox{.}}{2008}]%
        {MDeXploration}
\bibfield{author}{\bibinfo{person}{A. Simitsis}, \bibinfo{person}{A. Baid},
  \bibinfo{person}{Y. Sismanis}, {and} \bibinfo{person}{B. Reinwald}.}
  \bibinfo{year}{2008}\natexlab{}.
\newblock \showarticletitle{Multidimensional content eXploration}.
\newblock \bibinfo{journal}{\emph{PVLDB}} (\bibinfo{year}{2008}).
\newblock


\bibitem[\protect\citeauthoryear{Skovsgaard, Sidlauskas, and Jensen}{Skovsgaard
  et~al\mbox{.}}{2014}]%
        {TopTerms}
\bibfield{author}{\bibinfo{person}{A. Skovsgaard}, \bibinfo{person}{D.
  Sidlauskas}, {and} \bibinfo{person}{C.~S. Jensen}.}
  \bibinfo{year}{2014}\natexlab{}.
\newblock \showarticletitle{Scalable top-k spatio-temporal term querying}.
\newblock \bibinfo{journal}{\emph{ICDE}} (\bibinfo{year}{2014}).
\newblock


\bibitem[\protect\citeauthoryear{Walther and Kaisser}{Walther and
  Kaisser}{2013}]%
        {textStreams}
\bibfield{author}{\bibinfo{person}{M. Walther} {and} \bibinfo{person}{M.
  Kaisser}.} \bibinfo{year}{2013}\natexlab{}.
\newblock \showarticletitle{Geo-spatial Event Detection in Twitter Stream}.
\newblock \bibinfo{journal}{\emph{ECIR}} (\bibinfo{year}{2013}).
\newblock


\bibitem[\protect\citeauthoryear{Wang, Cao, and Yu}{Wang et~al\mbox{.}}{2020}]%
        {spatioTemporal}
\bibfield{author}{\bibinfo{person}{S. Wang}, \bibinfo{person}{J. Cao}, {and}
  \bibinfo{person}{P. Yu}.} \bibinfo{year}{2020}\natexlab{}.
\newblock \showarticletitle{Deep learning for spatio-temporal data mining: A
  survey}.
\newblock \bibinfo{journal}{\emph{TKDE}} (\bibinfo{year}{2020}).
\newblock


\bibitem[\protect\citeauthoryear{{Yu}, {Xu}, {Wang}, and {Ni}}{{Yu}
  et~al\mbox{.}}{2019}]%
        {TopicHierarchyModeling}
\bibfield{author}{\bibinfo{person}{D. {Yu}}, \bibinfo{person}{D. {Xu}},
  \bibinfo{person}{D. {Wang}}, {and} \bibinfo{person}{Z. {Ni}}.}
  \bibinfo{year}{2019}\natexlab{}.
\newblock \showarticletitle{Hierarchical Topic Modeling of Twitter Data for
  Online Analytical Processing}.
\newblock \bibinfo{journal}{\emph{IEEE Access}} (\bibinfo{year}{2019}).
\newblock


\bibitem[\protect\citeauthoryear{Zhang and Han}{Zhang and Han}{2019}]%
        {MMMTD}
\bibfield{author}{\bibinfo{person}{C. Zhang} {and} \bibinfo{person}{J. Han}.}
  \bibinfo{year}{2019}\natexlab{}.
\newblock \showarticletitle{Multidimensional Mining of Massive Text Data}.
\newblock \bibinfo{journal}{\emph{DMKD}} (\bibinfo{year}{2019}).
\newblock


\bibitem[\protect\citeauthoryear{Zhang, Zhai, Han, Srivastava, and Oza}{Zhang
  et~al\mbox{.}}{2009}]%
        {TopicCube}
\bibfield{author}{\bibinfo{person}{D. Zhang}, \bibinfo{person}{C.~X. Zhai},
  \bibinfo{person}{J. Han}, \bibinfo{person}{A. Srivastava}, {and}
  \bibinfo{person}{N. Oza}.} \bibinfo{year}{2009}\natexlab{}.
\newblock \showarticletitle{Topic Modeling for OLAP on Multidimensional Text
  Databases}.
\newblock \bibinfo{journal}{\emph{Stat. Anal. Data Min.}}
  (\bibinfo{year}{2009}).
\newblock


\bibitem[\protect\citeauthoryear{Zhao, Chen, and Cong}{Zhao
  et~al\mbox{.}}{2016}]%
        {TopicExploration}
\bibfield{author}{\bibinfo{person}{K. Zhao}, \bibinfo{person}{L. Chen}, {and}
  \bibinfo{person}{G. Cong}.} \bibinfo{year}{2016}\natexlab{}.
\newblock \showarticletitle{Topic Exploration in Spatio-Temporal Document
  Collections}.
\newblock \bibinfo{journal}{\emph{SIGMOD}} (\bibinfo{year}{2016}).
\newblock


\end{thebibliography}

\clearpage

\appendix

\section{STTOLAP Operators} \label{sec:sttolap}
A data cube allows different \textbf{O}n\textbf{L}ine \textbf{A}nalytical \textbf{P}rocessing (OLAP) operators to group, filter, and analyze cells and subsets of cells at different levels of granularity and under different perspectives. 
Those operators are known as \emph{Slice}, \emph{Dice}, \emph{Roll-Up}, and \emph{Drill-Down}~\cite{dwbook}, and they take as input a cube and produce as output another cube.
In the following, we extend the basic OLAP operators to \emph{STTOLAP operators}, i.e., for spatial, textual, and temporal dimensions, hierarchies, and measures.
In general, an OLAP (and STTOLAP) operator \textit{OP} accepts as input a cube $C'$, some parameters \textit{params} and outputs a new cube $C''$, i.e., \textit{OP}$(C'$, \textit{params}$)$$=$$C''$.
In this way, a new OLAP operator can be applied to $C''$.
Among all cubes, we distinguish the initial or \emph{base cube} $C$ as the cube containing all the original information at the base level.
In the following, we generally assume every OLAP operator \textit{OP} to have access to $C$ since some operators need access to the base cube $C$ to produce the desired result.
\subsection{STT-Slice}
Slice operates over the current data cube $C'$ and takes as a parameter a dimension member $v_i$ in a specific level $l_i$ and dimension $d_i$.
It keeps only cells in $C'$ corresponding to $v_i$ and finally removes dimension $d_i$.
An example of the slice operator is ``slice location dimension on user-defined polygon representing Aalborg''.

\para{STT-Slice.} operator is defined as \emph{STTSlice}$($$C_{stt}',v_{i}$$)$$=$$C_{stt}''$.
It takes an \textit{n}-dimensional \emph{STTCube} $C_{stt}'$ and a member $v_i$ of the level $l_i$ of the spatial, textual, or temporal dimension $d_{Location}$, $d_{Text}$, or $d_{Time}$, respectively, and produces a resulting cube $C_{stt}''$ with $n-1$ dimensions, i.e., it returns measure values aggregated at level $l_i$ only for the member $v_i$ and removes the respective dimension.
The member $v_i$ could be an object in the taxonomy of a semantic-based hierarchy (e.g., Aalborg as a dimension member in $l_{City}$) or could be a grid cell at some granularity level.
Similarly, $v_i$ could be a specific \textit{Theme} or \textit{Topic} for the textual dimension and a specific \emph{day} or \emph{month} for the temporal dimension.
%

\subsection{STT-Dice}
While the slice operator selects and removes a single dimension, the dice operator produces a new cube whose cell contents have been filtered based on a set of conditions (complex predicates, queries covering several cells) but without removing any dimension.
That is, it produces a resulting cube with the same number of dimensions but based on facts that satisfy the provided set of conditions.
Such conditions can use a combination of spatial, textual, temporal, and general-purpose functions.
These functions can perform different computations, e.g., they combine more than one object and return a new single aggregated object, or compare two objects and return a Boolean value, or they can produce a numeric value based on some computation.

\para{STT-Dice.}
The \textit{STT-Dice} operator selects only cell(s) that satisfy the provided spatial, textual, or temporal logical conditions.
Given a STTCube $C_{stt}'$ and a set of logical conditions \textit{COND}$_i$, the \textit{STTDice} is represented as
$STTDice(C_{stt}',\textit{COND}_i) = C_{stt}''$ where \textit{COND}$_i$ is a set of atomic or compound spatial, textual, or temporal logical conditions.
For instance, it can return only the cell(s) that intersect with the polygon describing a custom region of interest and containing at least $n$ observations, the cell containing only objects whose relevance score with the topic \emph{Food} is above $0.7$ and that contain at least $10$ terms, or the cells that discuss a specific topic in May 2020.


\subsection{STT-Roll-Up}
The Roll-Up operator aggregates measure values along a hierarchy by moving from a child level to a parent level. 
This allows to move the analysis to a coarser granularity.
An example of this operator is ``Roll-Up from City to Country''.

\para{STT-Roll-Up.}
The \textit{STT-Roll-Up} operator groups facts by aggregating the measure values of all child members that belong to the same parent member in a spatial, textual, or temporal hierarchy.
Given a STTCube $C_{stt}'$, a child level $l_{i\downarrow}$ and a target parent level $l_{i\uparrow}$ (identifying a hierarchy step function $hs_i$ in the spatial, textual, or temporal dimensions), the \textit{STTRollUp} operator defined as
\textit{STTRollUp}$(C_{stt}',l_{i\downarrow},l_{i\uparrow}) = C_{stt}''$
produces a new STTCube that has the same number of dimensions and for each measure $m{\in}M$, the aggregation function associated with $m$ is used to create an aggregated measure value for it at the higher parent level.
For instance, when we \textit{Roll-Up} from \textit{City} level to the \textit{Region} level, \textit{Fact Count} values get summed to compute the new total.
Similarly, ``Roll-Up to Topic level from Theme level'' groups facts by aggregating the measure values of all \emph{Themes} that belong to the same \emph{Topic} and ``Roll-Up to Quarter level from Month level'' groups facts by aggregating the measure values of all \emph{Months} that belong to the same \emph{Quarter}.
%

\subsection{STT-Drill-Down}
The inverse of the Roll-Up is the Drill-Down operator which shows data at finer granularity by dis-aggregating measure values along a hierarchy.
An example of the Drill-Down operator is ``Drill-Down to City level from Region level''.
For this operation, the base cube is required as we cannot uniquely dis-aggregate measure values knowing only the values at the parent level.

\para{STT-Drill-Down.}
Given an STTCube $C_{stt}'$ at a non-base level $l_{i\uparrow}$ for the spatial ($d_{Location}$), textual ($d_{Text}$), or temporal ($d_{Time}$) dimension and a target child-level $l_{i\downarrow}$ (identifying a hierarchy step function $hs_i$), the \textit{STTDrillDown} operator \textit{SDrillDown}$(C_{stt},C_{stt}',$ $l_{i\uparrow},l_{i\downarrow})$$=$$C_{stt}''$ produces a new STTCube at finer granularity computing dis-aggregated measure values along the selected spatial, textual, or temporal hierarchy, respectively.
For instance, \textit{STT-Drill-Down} can move from the \textit{Region} level to the \textit{City} level dis-aggregating the \textit{Fact Count} of each \textit{Region} in the counts for each \textit{City} in that \textit{Region}.
``Drill-Down to Term from Theme'' following the inverse textual hierarchy step \textit{Term}$\leftarrow$\textit{Theme} and ``Drill-Down to Day from Year'' following the inverse temporal hierarchy step \textit{Day}$\leftarrow$\textit{Year} are examples of STT-Drill-down operators for textual and temporal dimension, respectively.

\begin{figure}[t]
	\centering
	\begin{tikzpicture}[align=center,node distance=9cm, font=\scriptsize]

			\node at (0,3) (nodeDCT) {DLT (100M)({T})};
			\node at (-1,2) (nodeDC) [below left=0.4cm and 0.6cm of nodeDCT] {DL (15M)({F})};
			\node at (0,2) (nodeDT) [below=0.4cm of nodeDCT] {DT (4M)({T})};
			\node at (1,2) (nodeCT) [below right=0.4cm and 0.6cm of nodeDCT] {LT (96M)({F})};
			\node at (-1,1) (nodeD) [below=0.4cm of nodeDC] {D (37)({F})};
			\node at (0,1) (nodeC) [below=0.4cm of nodeDT] {L (14M)({F})};
			\node at (1,1) (nodeT) [below=0.4cm of nodeCT] {T (2M)({F})};
			\node at (0,0) (nodeNone) [below=0.4cm of nodeC] {~~$\star$ (1)({F})};

			\draw (nodeDCT) -- (nodeDC);
			\draw (nodeDCT) -- (nodeDT);
			\draw (nodeDCT) -- (nodeCT);

			\draw (nodeDC) -- (nodeD);
			\draw (nodeDC) -- (nodeC);
			\draw (nodeDT) -- (nodeD);
			\draw (nodeDT) -- (nodeT);
			\draw (nodeCT) -- (nodeC);
			\draw (nodeCT) -- (nodeT);

			\draw (nodeD) -- (nodeNone);
			\draw (nodeC) -- (nodeNone);
			\draw (nodeT) -- (nodeNone);
	\end{tikzpicture}
	\captionsetup{font=footnotesize, skip=0pt}
	\caption{Spatio-Textual-Temporal Lattice}
	\label{fig:latticeExample}
\end{figure}

\begin{figure*}[thb!]
	\centering
	\footnotesize

	\ref{QRTTrendLegendApp}
	\captionsetup{font=footnotesize, skip=0pt}
	\caption{QRT Vs Data Size}
	\label{fig:QRTsTrends_App}
\end{figure*}

\begin{figure*}[t]
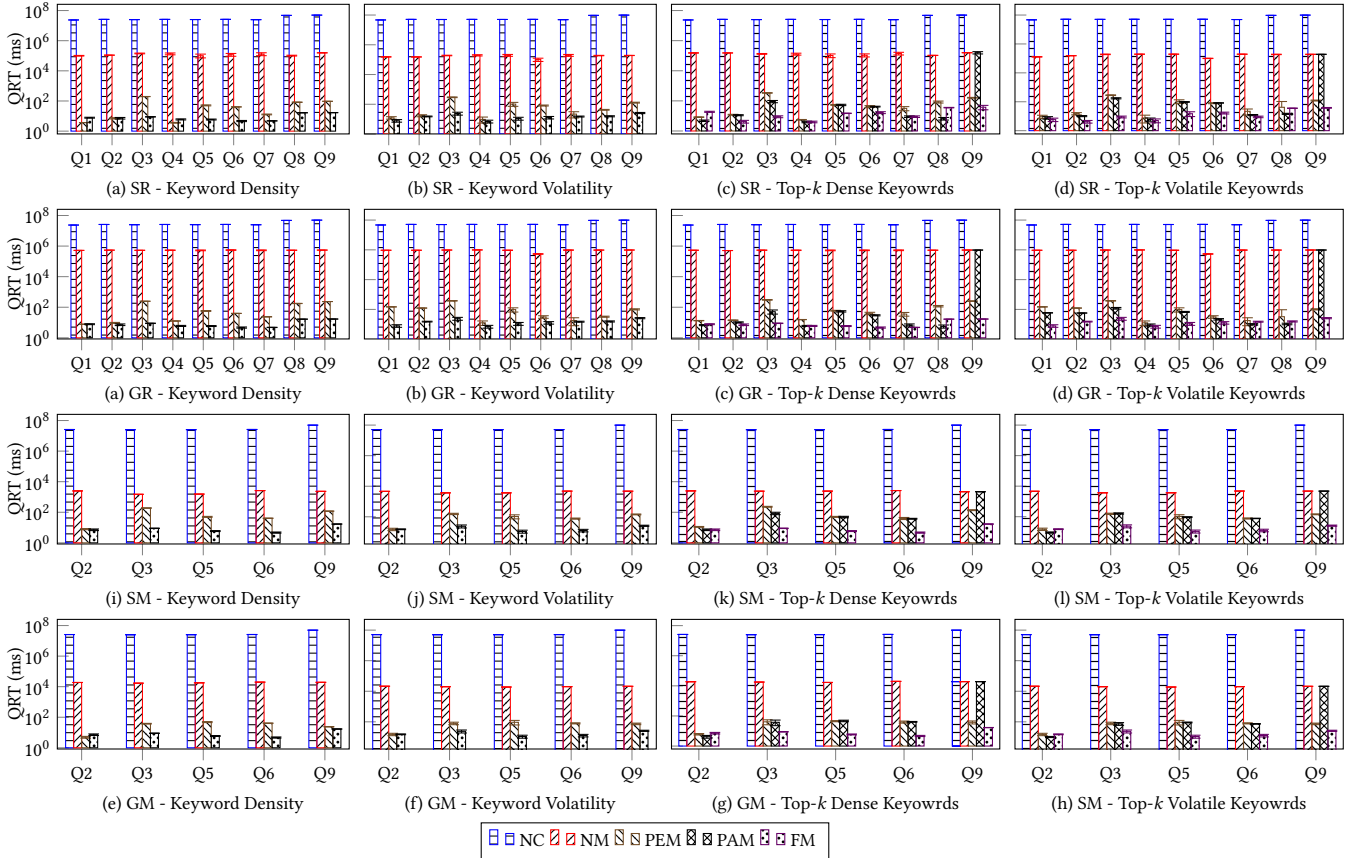

	\centering
	\footnotesize
	%
	\ref{namedApp}
	\captionsetup{font=footnotesize, skip=0pt}
	\caption{QRTs for spatio-textual-temporal measures for different combinations of hierarchy schemes over 125 Million of Data}
	\label{fig:QRT125M_App}
\end{figure*}

\section{Lattice Example} \label{sec:latticeExampleApp}

Consider a simple lattice of \emph{cuboid} for a cube with 3 dimensions (Figure~\ref{fig:latticeExample}), each with a single 2-level hierarchy, namely with base levels Location (L) with $14$M rows for the spatial dimension,  Term (T) with $2$M rows for the textual dimension, and Date (D) with $37$ rows for the temporal dimension, each of which can be then rolled up to the \emph{all} level ($\star$) with only one row.
Each node in the lattice is associated with two values, first, the \emph{number of rows} in the cuboid, and second a \emph{flag} (T/F) to mark if the current cuboid is materialized.
At the top of the lattice, we have the base cuboid (which is always materialized) with Date, Location, and Term (DLT) containing in this example all rows ($100$M).
If we Roll-Up the spatial dimension from Location to All we obtain a new cuboid (DT) with $4$M rows.
The cuboid DLT is referred to as the \textit{ancestor} of the cuboid DT.
If the cube is partially materialized, i.e., not all cuboids are materialized, and the cuboid DT is materialized, then to obtain the \textit{Fact Count} for every Date and Term, the cuboid DT with $4$M rows would contain the answer already pre-computed without the need to compute such an answer from the base cube DLT with $100$M rows.
Moreover, when the cuboid T is not materialized, we can still compute the \textit{Fact Count} for every Term from the cuboid DT by accessing only 4M values instead of the $100$M in DLT.

\section{Cost Model} \label{sec:costmodelApp}

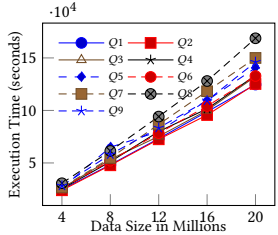
\begin{figure}[t]
\centering
\begin{tikzpicture}
\scriptsize
\begin{axis}[
xlabel={Data Size in Millions},
ylabel={Execution Time (seconds)},
legend pos= north west,
legend columns = -1,
legend columns=2,
height=4cm,
legend style={nodes={scale=0.7},fill=none,draw=none},
y tick label style={scaled y ticks=base 10:-4},
xlabel style={yshift = {5}},
ylabel style={yshift = {-5}},
xtick pos=left,
ytick pos=left,
xtick=data,
]

\addplot coordinates {(4,24648.70)(8,48260.10)(12,73213.10)(16,98755.40)(20,124648.50)};
\addplot coordinates {(4,23967.40)(8,47751.20)(12,72497.30)(16,95630.40)(20,125171.40)};
\addplot +[mark=triangle] coordinates {(4,25986.90)(8,53469.70)(12,75315.20)(16,100062.90)(20,132325.30)};
\addplot coordinates {(4,25316.50)(8,51130.60)(12,79952.00)(16,101543.20)(20,133375.50)};
\addplot +[dashed] coordinates {(4,26250.30)(8,64949.40)(12,79897.50)(16,109528.00)(20,140530.50)};
\addplot coordinates {(4,25622.60)(8,52111.30)(12,78329.00)(16,104110.10)(20,132751.00)};
\addplot coordinates {(4,26601.70)(8,54698.80)(12,86688.60)(16,118287.80)(20,149966.20)};
\addplot coordinates {(4,30670.40)(8,61393.00)(12,94152.30)(16,128014.50)(20,168694.59)};
\addplot coordinates {(4,28257.20)(8,58148.40)(12,83420.40)(16,109384.20)(20,146084.00)};

\legend{$Q1$,$Q2$,$Q3$,$Q4$,$Q5$,$Q6$,$Q7$,$Q8$,$Q9$}
\end{axis}
\end{tikzpicture}
\captionsetup{font=footnotesize, skip=0pt}
\caption{QRT Vs Data Size}
\label{fig:QRTVsDS}
\end{figure}

The core of the proposed \emph{partial materialization} approach depends on the trade-off between the storage cost of materializing any particular cuboid and the actual \emph{benefit} that the materialization of the cuboid provides.
To evaluate this benefit, we have to estimate the (run time) cost of a query.
To devise a cost model for this estimation, we performed a micro-benchmark which confirmed that the running time is directly proportional to the data size (the number of rows). Hence we can use the following linear cost model for benefit calculation
we selected a set of representative queries (Q1---Q9, details in Table~\ref{tab:queries}) for the aforementioned \emph{spatio-textual-temporal measures} and measured the runtime of these queries on increasing data sizes.
The micro-benchmark (Figure~\ref{fig:QRTVsDS}) confirmed that the running time is directly proportional to the data size (the number of rows), i.e., it confirms that we can use the Linear Cost Model~\cite{ImpCubeEff} and the associated benefit calculation.
Then, to model the dependency relationships among all the possible cuboids we use the lattice framework~\cite{ImpCubeEff} (Figure~\ref{fig:latticeExample}).
Hence, to compute the benefit of materializing a particular cuboid $c$, we need to compare the cost of answering queries at all levels of granularity (i.e., for the current cuboid $c$ and all its descendants in the lattice) with the current set of materialized cuboids against the cost when $c$ is also materialized.
$$\textit{Benefit}(c) = \sum_{c' \in \textit{descendants}(c) \cup \{c\}} \textit{cost}(c')-\textit{size}(c)$$

For instance, assume the lattice in Figure~\ref{fig:latticeExample} and that only the base cuboid (DLT) with $100$M rows is materialized.
If we consider DT with $4$M rows, we have that, if materialized, queries against the cuboids D, T, $\star$, and DT itself can be answered through it (with a cost of $4$M), while without materializing DT, we will need to compute the answer against DLT (with a cost of $100$M).
Hence, materializing DT will achieve a benefit of $(100 - 4) * 4 = 384$M.
Whereas, materializing LT with $96$M rows does not achieve a significant benefit ($(100 - 96) * 4 = 16$M).
\end{document}